\documentclass[a4paper,fleqn,usenatbib,usedcolumn]{mnras}

\usepackage[T1]{fontenc}
\usepackage{ae,aecompl}

\usepackage{graphicx}	
\usepackage{amsmath}	
\usepackage{amssymb}	

\newcommand{\msol}{\mathrm{M}_{\sun}}					
\newcommand{\usfr}{\mathrm{M}_{\sun}~\mathrm{yr}^{-1}}			
\newcommand{\cgsdens}{\mathrm{g~cm}^{-3}}				
\newcommand{\vect}[1]{\mathbfit{#1}}  
\newcommand{\flash}{\textsc{flash}}

\newcommand{\smallsub}[1]{\mathrm{#1}}

\defcitealias{derijcke04}{De Rijcke et al. (2004)}
\defcitealias{vdbosch08}{Van den Bosch et al. (2008)}

\title[Refueled and shielded]{Refueled and shielded - The early evolution of Tidal Dwarf Galaxies}
\author[B. Baumschlager et al.]{
Bernhard Baumschlager,$^{1,2}$\thanks{E-mail: bernhard.baumschlager@astro.uio.no}
Gerhard Hensler,$^{1}$
Patrick Steyrleithner$^{1}$ 
and 
\newauthor{ Simone Recchi$^{1}$}
\\
$^{1}$Department of Astrophysics, University of Vienna, T\"urkenschanzstrasse 17, 1180 Vienna, Austria\\
$^{2}$Institute of Theoretical Astrophysics, University of Oslo, Postboks 1029, Blindern, 0315 Oslo, Norway
}

\date{Accepted XXX. Received YYY; in original form ZZZ}
\pubyear{2018}

\begin{document}
\label{firstpage}
\pagerange{\pageref{firstpage}--\pageref{lastpage}}
\maketitle

\begin{abstract}
We present, for the first time, numerical simulations of young tidal dwarf galaxies (TDGs), including a self-consistent treatment of the tidal arm in which they are embedded. 
Thereby, we do not rely on idealised initial conditions, as the initial data of the presented simulation emerge from a galaxy interaction simulation.
By comparing models which are either embedded in or isolated form the tidal arm, we demonstrate its importance on the evolution of TDGs, 
as additional source of gas which can be accreted and is available for subsequent conversion into stars.
During the initial collapse of the proto-TDG, with a duration of a few 100 Myr, the evolution of the embedded and isolated TDGs are indistinguishable.
Significant differences appear after the collapse has halted and the further evolution is dominated by the possible accretion of material form the surroundings of the TDGs.
The inclusion of the tidal arm in the simulation of TDGs results in roughly a doubling of the gas mass ($M_\mathrm{gas}$) and gas fraction ($f_\mathrm{gas}$), 
an increase in stellar mass by a factor of 1.5 and a $\sim3$ times higher star formation rate (SFR) compared to the isolated case. 
Moreover, we perform a parametric study on the influence of different environmental effects, i.e. the tidal field and ram pressure.
Due to the orbit of the chosen initial conditions, no clear impact of the environmental effects on the evolution of TDG candidates can be found.
\end{abstract}

\begin{keywords}
methods: numerical -- hydrodynamics -- galaxies: dwarf -- galaxies: evolution
\end{keywords}



\section{Introduction}\label{sec:intro}		

During gravitational interactions of galaxies, long filamentary arms are drawn out of the galaxies main bodies by tidal forces \citep[e.g.][]{toomre72}.
Within these so-called tidal arms or tails accumulations of gas and stars can be found.
These clumps are also detectable through their H$\alpha$ and FUV emission, tracers of recent or ongoing star formation (SF), 
and harbour molecular gas, the reservoir for SF \citep[e.g.][]{mirabel91,mirabel92,braine00,braine01,braine04}.
If these objects are massive enough, i.e. masses above $10^8~\msol$ \citep{duc12}, they are commonly referred to as tidal dwarf galaxies (TDGs).

Due to the formation out of tidally stripped material and their shallow potential wells, TDGs are expected to be almost entirely free of dark matter (DM), 
requiring mass-to-light ratios close to one \citep[e.g.][]{barnes92,wetzstein07}, in agreement with observations \citep[e.g.][]{duc94}.
Nevertheless, TDGs are sharing several properties with primordial dwarf galaxies (DGs), i.e. those formed in low mass DM halos in the standard model of cosmology.

Despite their different formation mechanism, TDGs show similar sizes, masses, and properties as normal primordial DGs, 
with effective radii in the order of 1 kpc and masses of the stellar content between $10^6 \le \mathrm{M}_\star/\msol \le 10^{10}$ 
and therefore follow the same mass-radius (MR) relation \citep[][and references therein]{dabringhausen13}.

The luminosity range typical for observed TDGs is $-14<M_B<-18$ \citep[e.g.][]{mirabel92,duc94,weilbacher03}, comparable to the primordial DG population \citep[e.g.][]{mo10}.

As TDGs form out of preprocessed material expelled from their parent galaxies, they are expected to deviate from the mass-metallicity relation of DGs, 
being more metal-rich then primordial DGs of the same mass \citep{duc94,croxall09,sweet14}.
However, \citet{recchi15} showed that this is only true for recently formed TDGs, because they are formed out of substantially pre-enriched material. 
This can als mean that the mass-metallicity relation could be an effect of different TDG-formation times. 
TDGs formed in the early universe, emerging from relatively metal-poor material, could constitute the low-metallicity end of this relation, 
as self-enrichment is the dominant process in their chemical evolution.

Many TDGs show a bimodal distribution in the ages of their stellar populations, a combination of old stars expelled from the parent galaxies and a young in-situ formed population.
Thereby, not more than 50\% of the stars in a TDG emerge from the parent galaxies to constitute the old population.
The majority of the young population of stars is born during a starburst at the time the TDG is formed.
Further SF occurs with star-formation rates (SFRs) as expected for DGs of comparable size \citep[e.g.][]{elmegreen93,hunter00}.
Observationally derived SFRs of TDGs range from $10^{-4}$ to $10^{-1}~\usfr$ \citep[e.g][]{duc98,leewaddell14,lisenfeld16}.

Up to now many observations established the existence of TDGs and their properties but there still remain unanswered questions, such as: 
How do TDGs form within the tidal arms?
What is the production rate of TDGs? 
How many of them survive disruptive events, like stellar feedback and the parent galaxies' tidal field? 
How many of the TDGs formed during a galaxy interaction fall-back to the parent galaxies, dissolve to the field or become satellite galaxies of the merger remnant?
What is the contribution of TDGs to the present day DG population?
These questions can typically be addressed by numerical simulation only.

The first galaxy interaction simulation which showed the possibility of DG formation within tidal arms was performed by \citet{barnes92}.
The TDGs produced in these N-body/SPH simulations were formed by the accumulation of stars which form a gravitational bound structure and bind gas to them, 
whereas in the simulations of \citet{elmegreen93}, the gas within the tidal arm forms Jeans unstable clouds which collapse and enables the formation star.

\citet{wetzstein07} investigated numerical resolution effects on the formation of TDGs and concluded that, within high resolution N-body/SPH simulations, 
the presence of a dissipative component, i.e. gas, is required in order to form TDGs, but also that the spatial extend of the gaseous disk has to be large enough.

Due to the absence of a supporting DM halo, TDGs are suspected to be very vulnerable to internal and external disruptive effects like stellar feedback, 
ram pressure or a tidal field \citep[e.g.][]{bournaud10}.
But both observations and simulation have shown that TDGs can indeed be long-lived objects and are able to survive for several billion years.

Within a large set of galaxy merger simulation \citet{bournaud06} found that half of the formed TDGs with masses greater than $10^8~\msol$ can survive for a Hubble time.
\citet{recchi07} performed 2D simulations of TDGs and found that these objects can survive an initial starburst for several 100 Myr and might turn into long-lived dwarf spheroidal galaxies.
Already \citet{hensler04} have demonstrated by means of 1D simulations which are more vulnerable to supernova explosions that DGs remain bound even without the gravitational aid of a DM halo 
but become gas-free similar to dwarf spheroidal galaxies (dSph).

The potentially devastating effect of ram-pressure stripping (RPS) on isolated TDGs has been investigated by \citet{smith13}.
Thereby, they found that for wind speeds in excess of $200\,\mathrm{km\,s}^{-1}$ even the stellar disk of a TDG can be heavily disrupted due to the loss of gas caused by RPS.
As the gas gets removed from the TDGs disk, the gravitational potential gets weakened and more and more stars become unbound.

Observationally, the oldest confirmed TDGs have been found by \citet{duc14} with approximate ages of about 4 billion years.
More recent detailed chemodynamical simulations on the survivability of TDGs have shown that, despite their lack of a supporting DM halo, 
TDGs can survive disruptive events like starbursts or the tidal field of their parent galaxies and can reach ages of more than 3 Gyr \citep{ploe14,ploe15}.

Recently, \citet{ploe18} demonstrated the possibility of identifying TDGs within the high-resolution cosmological simulation runs of the EAGLE suite \citep{schaye15}, 
without any further conclusions about the TDG formation rate or survivability, due to the limited temporal and spatial resolution.

The fraction of TDGs among the total present-day DG population is estimated to $5-10\%$, 
depending on the detailed assumption of the TDG production and survival rates \citep[e.g.][]{bournaud10,wen12,kaviraj12}.
The most extreme estimates, considering higher production rates during gas-rich merges events in the early universe \citep{okazaki00}, 
or based on the structural similarities of TDGs and DGs \citep{dabringhausen13} predict that the whole DG population is of tidal origin.

All the previous theoretical and numerical work on TDGs either focuses on the formation during galaxy interactions, their production rate and long-term survivability, 
or the evolution of TDGs as individual objects decoupled from the tidal tail. 
In all these studies the influence of the tidal tail as a dynamically developing live environment itself is neglected or it has not been taken into account at first place.
In contrast to this, we include for the first time a self-consistent treatment of the tidal arm in detailed numerical studies of TDGs,
in order to study TDGs in the dynamical course of the tidal arms.

This paper is structured as followed. In Section \ref{sec:code}, a short overview of the simulation code and the implemented physical processes is provided.
Section \ref{sec:model} describes the different models and the setup of the simulation runs.
The results are presented in Section \ref{sec:results} and a final discussion is given in Section \ref{sec:discussion}.

\section{The Code}\label{sec:code}		

The simulations presented in this work make use of an extended version of the Adaptive Mesh-Refinement (AMR) code \flash\,3.3 \citep{fryxell00}, 
applying the Monotonic Upstream-centred Scheme for Conservation Laws (MUSCL) Hancock scheme in combination with the HLLC Riemann solver and van Leer slope limiter for hydrodynamics.
The simulations are advanced on the minimal timestep according to the CFL condition \citep{courant67}, throughout this work the CFL constant is set to 0.1 for stability reasons.
In simulations including the effect of RPS, the high density and velocity contrast at the interface between tidal arm and circumgalactic medium (CGM) can lead to numerical instabilities resulting in unphysical negative densities. 
In addition, timescales of cooling and SF are temporarily shorter than the dynamical timescale and, by this, these processes are insufficiently resolved in time. 
All these pitfalls are avoided by choosing a lower value for the CFL factor than in pure hydrodynamics, but make the numerical modeling time expensive.
Self-gravity within the simulation box is solved using the Multigrid Poisson solver \citep{ricker08} provided by \flash.

The code was extended by \citet{ploe14,ploe15} in order to account for a metal dependent cooling of gas, SF, stellar feedback, a tidal field and RPS.
A short overview of the extensions and their models is provided below, while for a more detailed description the reader is referred to \citet{ploe14,ploe15}.

\subsection{Cooling}\label{sub:cool}			
In order to account for a composition dependent radiative cooling the abundances of H, He, C, N, O, Ne, Mg, Si, S and Fe of each grid cell are traced.
The cooling function is split into two different temperature regimes at $10^4 \,\mathrm{K}$.
For gas temperatures above $10^4\,\mathrm{K}$ cooling functions from \citet{boehringer89} for an optically thin plasma in collisional ionisation equilibrium,
taking H, He, C, N, O, Ne, Mg, Si, S and Fe into account, are used.
At temperatures below $10^4\,\mathrm{K}$ the cooling functions from \citet{schure09,dalgarno72} are evaluated for C, N, O, Si, S and Fe.
The energy loss due to radiative cooling is calculated at every time-step solving an implicit equation by means of the Newton-Raphson method.

\subsection{Stars}\label{sub:stars}			
Within the presented simulations stars are represented by particles, which are advanced in the potential of the TDG by a variable-timestep leapfrog algorithm.
Thereby one stellar particle represents a whole star cluster, instead of a single star, with its underlying stellar population described by an invariant initial mass function (IMF).

\subsubsection{Star formation}\label{sub:sf}			
The SF is fully self-regulated following the description of \citet{kth95}, where the SFR is determined by the stellar birth function:
\begin{equation}\label{equ:sbf}
  \psi(\rho,\mathrm{T}) = C_2\, \rho^2\, e^{T/\mathrm{T}_\mathrm{s}},
\end{equation}
where $\rho$ and $T$ are the gas density and temperature, $\mathrm{T}_\mathrm{s}~=~1000\,\mathrm{K}$ and $C_2=2.575\times10^8 \,\mathrm{cm}^3\,\mathrm{g}^{-1}\,\mathrm{s}^{-1}$.
Thereby, the exponential term acts as a temperature dependent efficiency function which allows for a smooth transition between a high SFR at low temperatures
and a low SFR at temperatures above $10^4\,\mathrm{K}$.

This formulation only depends on the local properties of a grid cell, i.e. the gas density and temperature and was derived as self-regulated recipe. Therefore, this approach does not require any global scaling relation and was successfully applied to massive spirals by \citet{samland97} and to dwarf irregular galaxies by \citet{hensler02} leading to realistic SFRs and chemical abundances. 
In parallel studies we compare this self-regulation recipe with that of density and temperature thresholds as described by e.g. \citet{stinson06} (Keuhtreiber et al. in preparation), and in others with respect to the initial mass function at low SFRs \citep[][Steyrleithner et al. in preparation]{hensler17}.

In principle Equation~\ref{equ:sbf} allows for very low SFRs in hot and low density regions, which would result in large numbers of very low mass star clusters.
To avoid these low mass stellar particles a threshold on the stellar birth function is applied in the form of
\begin{equation}\label{equ:sbf_thresh}
 \psi_\mathrm{thresh} = \theta_\mathrm{sf} \; \frac{3\,M_\mathrm{min}}{4\,\pi\,r^3_{\smallsub{GMC}}\,\tau_\mathrm{cl}},
\end{equation}
where $M_\mathrm{min}=100\,\msol$ is the minimal cluster mass, $\tau_\mathrm{cl}~=~1\,\mathrm{Myr}$ the cluster formation time, 
$r_{\smallsub{GMC}}=160\,\mathrm{pc}$ the size of a giant molecular cloud and $\theta_\mathrm{sf}$ is a dimensionless factor.
For $\theta_\mathrm{sf}=1$ - what is used here - and assuming a constant density and temperature during $\tau_\mathrm{cl}$, this threshold translates into a requirement on the minimal star cluster mass,
i.e. $M_\mathrm{min}=100\,\msol$ for the chosen parameters.

Whenever the SF criterion, in form of Equations~\ref{equ:sbf} and \ref{equ:sbf_thresh}, 
is fulfilled and no other actively star-forming particle is located within $r_{\smallsub{GMC}}$ a new star cluster is spawned.
From this time on a star cluster accumulates material from its surrounding according to Equation~\ref{equ:sbf}, 
until either the formation time $\tau_\mathrm{cl}$ expires or the IMF is filled.
As soon as a cluster is completed stellar feedback sets in, influencing the nearby grid cells.
If the SF criteria are still fulfilled within a grid cell, after an already existing cluster was closed for SF, a new particle is created within $r_{\smallsub{GMC}}$, 
while the feedback from the existing cluster already influences the stellar birth function as the temperature within a grid cell is increasing.

\subsubsection{Initial Mass Function}\label{sub:imf}			
The underlying stellar population of one stellar particle is described by an invariant multi-part power-law IMF \citep{kroupa01}
\begin{equation}\label{equ:imf}
  \xi(m) = k\; m^{-\alpha}
\end{equation}
where $k$ is a normalisation constant, $m$ the mass of a star and $\alpha$ the mass dependent power-law slope of the IMF
\begin{equation}\label{equ:alphaimf}
  \alpha = \begin{cases}
            0.3 ~~~\dots~~~ 0.01 \le m/\msol < 0.08 \\
            1.3 ~~~\dots~~~ 0.08 \le m/\msol < 0.5  \\
            2.3 ~~~\dots~~~ 0.5 \le m/\msol < 100.
           \end{cases}
\end{equation}
The number of stars $N$ and the mass $M_\mathrm{cl}$ of a star cluster can be calculated by:
\begin{equation}\label{equ:nstarimf}
  N = \int\limits_{m_\mathrm{min}}^{m_\mathrm{max}} \xi(m)\; dm
\end{equation}
\begin{equation}\label{equ:mclimf}
  M_\mathrm{cl} = \int\limits_{m_\mathrm{min}}^{m_\mathrm{max}} m\; \xi(m)\; dm
\end{equation}
For the presented simulations the IMF is split into 64 equally-spaced logarithmic mass bins in the mass range $0.1 \le m/\msol \le 120$.
All stars within one mass bin are represented by the average mass of the corresponding mass bin. 
The IMF of all star clusters, regardless of the cluster mass, is always filled to the highest mass bin, even with unphysical fractions of high-mass stars.
This treatment of the IMF in combination with the short cluster formation time resembles the maximum feedback case of \citet{ploe15}.

\subsubsection{Stellar feedback}		

For stars more massive than $8~\msol$ the feedback of type II supernovae (SNII), stellar winds and ionizing UV radiation is considered.
Below $8~\msol$ the release of energy and metals from type Ia supernovae (SNIa) and the enrichment of the interstellar medium (ISM) by Asymptotic Giant Branch (AGB) stars are included.

The different feedback mechanisms are considered at the end of a stars' lifetimes, 
except for ionizing UV radiation emitted by massive stars, which is continuously calculated at every timestep as long as a star cluster hosts stars above $8~\msol$.
The released energy of all these processes is injected into the gas as thermal energy.
The metallicity-dependent lifetimes of stars are taken from \citet{portinari98} and the stellar yields are combined from \citet{marigo96}, 
\citet{portinari98} and for SNIa from \citet[W7 model]{travaglio04}.

\paragraph{Stellar Wind:}			
During their evolution, high-mass stars lose a fraction of their mass due to radiation-driven winds, which heat and enrich the surrounding ISM.
The metal-dependent mass-loss rate due to stellar winds of OB star is given by \citep{hensler87,theis92}
\begin{equation}\label{equ:masslosswind}
 \dot{m} = 10^{-15}\, \left( \frac{Z}{\mathrm{Z}_\odot} \right)^{1/2} \left( \frac{L}{\mathrm{L}_\odot}\right)^{1.6}~\usfr,
\end{equation}
where $L$ is the stellar luminosity, derived from the mass-luminosity relation of \citet{maeder96}.
The heating of the ISM by stellar winds is determined by the winds' kinetic power
\begin{equation}\label{equ:windpower}
 \dot{E}_\mathrm{kin} = \frac{1}{2}\,\dot{m}\,v^2_\infty,
\end{equation}
with the stellar mass ($m$) dependent final wind velocity
\begin{equation}\label{equ:windvelocity}
 v_\infty = 3\times10^3 \left( \frac{m}{\msol} \right)^{0.15} \left( \frac{Z}{\mathrm{Z}_\odot}\right)^{0.08}\, .
\end{equation}
The total wind energy is then given by
\begin{equation}\label{equ:windenergy}
 E_\mathrm{wind}= \epsilon_\mathrm{wind}\, \dot{E}_\mathrm{kin}\, \tau_\mathrm{cl}\, N_\star\, ,
\end{equation}
where $\epsilon_\mathrm{wind}=5\%$ is the wind efficiency, $\tau_\mathrm{cl}$ is the age of a stellar particle and $N_\star$ is the number of stars in a mass-bin.
The kinetic heating of the ISM by stellar winds is typically one to two orders of magnitude lower compared to the heating by ionizing UV radiation, 
therefore it is only considered at the end of a stars lifetime.
The additional energy injected into the ISM by UV radiation of massive stars is treated separately.

\paragraph{Stellar Radiation:}			
Although the applied recipe of SF by \citet{kth95} already includes the self-regulation by stellar energy feedback, in recent simulations \citet{ploe15} we have implemented HII~regions as additional effect 
on SF self-regulation.
This recipe can also be used for momentum transfer and radiation-pressure driven turbulence \citep{renaud13}.
On galaxy scales this treatment (as a similar feedback scheme like ours) also produces galactic outflows \citep{bournaud14}.

The ionizing UV radiation of massive stars ($m\ge8~\msol$) is considered throughout their lifetime.
Therefore, it is treated separately from the winds driven by these stars.
Using the approach of a Str\"omgren sphere, the mass of ionised H around a single star in dependency of the stellar mass is calculated by
\begin{equation}\label{equ:mhii}
 M_\mathrm{\ion{H}{II}}\left( m \right) = \frac{S_\star \left( m \right) \mu^2 \, m_\mathrm{H}^2}{\beta_2 \, \rho \, f_\mathrm{H}^2}\, ,
\end{equation}
where $\rho$ and $\mu$ are the density and mean molecular weight of the ambient ISM,
$m_\mathrm{H}$ is the atomic mass of hydrogen, $\beta_2$ is the recombination coefficient, $f_\mathrm{H}$ is the mass fraction of H in the ISM
and the ionizing photon flux is derived as \citep{ploe15}
\begin{equation}\label{equ:sstar}
 S_\star = 3.6\times10^{42} \left( \frac{m}{\msol} \right)^4.
\end{equation}
The total mass of hydrogen ionized by one stellar particle is obtained by summing over all Str\"omgren spheres of the underlying high-mass population.
The temperature of the St\"omgren sphere is set to $2\times10^4\,\mathrm{K}$
and the temperature of a grid cell is calculated as mass-weighted average of all Str\"omgren sphere temperature and the temperature of the remaining mass in the grid cell.
Furthermore, the amount of ionized gas within a grid cell is tracked and excluded from SF until the last star more massive than 8~$\msol$ contained in a grid cell exploded as SNII. 
This ansatz is aimed only at reducing the cold gas mass that is capable for SF.
Since the Str\"omgren equation inherently includes the heating
and cooling balance, neither its spatial resolution nor its
contribution to the gas cooling must be considered.

\paragraph{Type II Supernovae:}			
When a star more massive than $8~\msol$ ends its life it will explode as core-collapse supernova, thereby heating the surrounding ISM and enriches it with heavy elements.
The energy injected into the ISM, by the stars inhabiting one IMF mass bin, is given by
\begin{equation}\label{equ:snii}
E_\mathrm{SNII,tot}=\epsilon_\mathrm{SNII}\, E_\mathrm{SNII}\,N_\star,
\end{equation}
where $\epsilon_\mathrm{SNII}$ is the SNII efficiency, $E_\mathrm{SNII}=10^{51}\,\mathrm{erg}$ the energy of one SNII, $N_\star$ the number of stars in one mass bin. 
Due to the lack of a value reliably derived on theoretical basis or from numerical studies \citep{thornton98} simplified 1D supernova explosion models) 
and although it must depend on the environmental state and therefore vary temporarily \citep{recchi14} $\epsilon_\mathrm{SNII}$ is set to 5\%.
The mass of the remaining neutron star or black hole is locked in the stellar particle and is not available for further SF.
The remnant mass for stars in the mass range $8\le \mathrm{m}/\msol\le120$ with solar metallicities range between $1.3$ and $2.1\,\msol$.

\paragraph{Type Ia Supernovae:}			
Based on the SNIa rate equations of \citet{matteucci01} and \citet{recchi09}
the probability of a star to be the secondary in a binary system is calculated as \citep{ploePhD}
\begin{equation}\label{equ:propsnia}
 P(m_2) = 2^{1+\gamma} \left( 1+\gamma \right) A \sum\limits_{i=i(m_2)}^\mathrm{nimf} \frac{\xi(m_i)}{\xi(m_2)} \mu_i^\gamma \Delta\mu \,,
\end{equation}
where $A=0.09$ is a normalisation constant, $\mu_i = m_2 / (m_2+m_i)$ is the binary mass ratio, $\xi$ is the IMF for the stellar masses $m_2$ and $m_i$ both in the mass range of $1.4-8~\msol$.
The distribution function of binary masses
\begin{equation}\label{equ:binary}
 f(\mu) = 2^{1+\gamma} \left( 1+\gamma \right) \mu^\gamma
\end{equation}
with $\gamma=2$, favouring equal mass systems, is incorporated in Equation \ref{equ:propsnia}.
The number of SNIa per IMF mass bin is then $N_\mathrm{SNIa} = P(m_2)\,N_\star$ and each SN is injecting $10^{51}~\mathrm{erg}$ with an efficiency of $\epsilon_\mathrm{SNIa}=5\%$ into the ISM.
As the main source of iron, SNIa also contribute significantly to the chemical evolution of a galaxy. 

\paragraph{Asymptotic Giant Branch Stars:}	
During the AGB phase stars suffer a strong mass loss, this is accounted for by the release of metals back to the ISM.
This enrichment of the ISM is considered at the end of an AGB-star's lifetime.
However, the energy injected into the ISM is neglected for this kind of stars. 
According to the treatment of SNIa a fraction of these stars will eventually end their life as SNIa.

\subsection{Environmental effects}		
In contrast to isolated DGs, TDGs form and evolve in the vicinity of their interacting parent galaxies, therefore, they feel their gravitational potential and the resulting tidal forces.
As the tidally stripped material expands and orbits around the host galaxy, it moves relative to the 
CGM 
and is therefore exposed to ram pressure.
For simplicity the CGM is assumed at rest with respect to the merger.

\subsubsection{Tidal field}			
Although the mass distribution of merging galaxies varies with time, a time constant gravitational potential of a point mass is assumed for the representation of the interacting host galaxies.
As the simulations are carried out in the rest frame of the TDG, the simulation box moves along an orbit around the interacting galaxies.
Therefore, the net-effect of the tidal field ($\vect{a}_\mathrm{tidal}$) on a given grid cell is calculated as the difference between the gravitational acceleration 
exerted by the external potential onto the grid cell ($\vect{g}_\mathrm{cell}$) and the acceleration felt by the simulation box ($\vect{g}_\mathrm{SB}$), i.e.
\begin{equation}\label{equ:tidal}
 \vect{a}_\mathrm{tidal} = \vect{g}_\mathrm{cell} - \vect{g}_\mathrm{SB}.
\end{equation}

\subsubsection{Ram Pressure}			
The simulations are carried out in the rest frame of the TDG, therefore a wind tunnel-like configuration is present.
In this setup the TDG stays at rest and the ambient CGM flows around it.
The resulting relative velocity of the CGM equals the negative orbital velocity of the TDG, e.g. $v_\mathrm{rp}=-v_\mathrm{orb}$, and is updated at every timestep.
The velocity of a grid cell is only modified to account for the varying velocity of the ambient medium when the fraction of material originating from the tidal arm is less then $10^{-4}$.

\section{The Models}\label{sec:model}		

\subsection{Initial Conditions}			
The initial conditions for the presented simulations emerge form a series of galaxy interaction simulations, which are presented in \citet{hammer10,fouquet12,hammer13,yang14}.
Already 1.5 Gyr after the first pericentre passage, the time at which the initial data are extracted, numerous forming TDG candidates can be identified.

\begin{figure}
 \centerline{\includegraphics[width=\columnwidth]{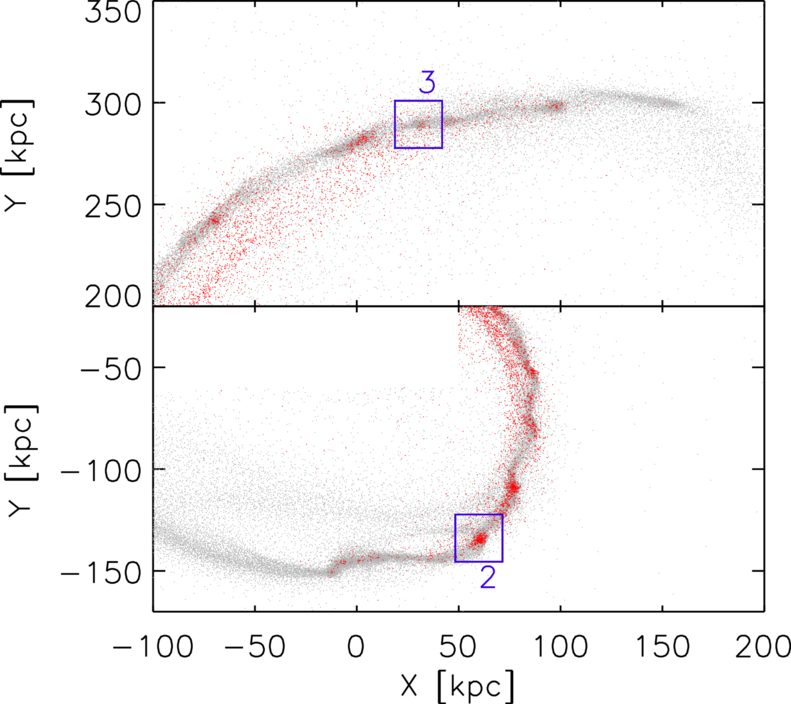}}
 \caption{Snapshot of the tidal arms 1.5 Gyr after the first pericentre passage in the \citet{fouquet12} simulation.
 For the purpose of demonstration only every 10th gas particle is shown as grey dot, the red points indicate the locations of stars expelled from the interacting galaxies.
 The blue boxes ehighlight the candidates and there later simulation boxes selected for this work.}
 \label{fig:initial}
\end{figure}

Out of the 15 TDG candidates of these simulation, we present follow up zoom-in simulations of two of the TDG candidates, 
which are highlighted by blue boxes and denoted by 2 and 3 in Figure~\ref{fig:initial}. 
These TDG candidates were already included in a comparative study of all cadidates by \citet{ploe15} with respect to their orientation and kinematics and were denoted there as TDG-c (3) and TDG-k (2).

\subsection{Data extraction}

Out of the many possible TDG candidates of the original \citet{fouquet12} simulation the candidate to test is selected according to the following criteria:
(1) An apparent over-density in the gaseous and stellar component along the tidal arm.
(2) A minimal gas mass within 2.5 kpc around the centre of the TDG of $M_\mathrm{g,min} \ga 10^8~\msol$.
(3) No other candidate within a distance of 15 kpc from the TDGs centre, to ensure that no overlap of the edge of the simulation box with a neighbouring TDG candidate can occur.

The mass centre of the TDG candidate is calculated iteratively. Starting from an initial guess of its position the mass centre within a sphere with $r_\smallsub{TDG}=4\,\mathrm{kpc}$ calculated by
\begin{equation}\label{equ:tdgcentre}
  \vect{R}_\mathrm{M} = \frac{1}{\sum\limits_i m_i}\sum\limits_i m_i\, \vect{r}_i,
\end{equation}
where $\vect{R}_\mathrm{M}$ is the mass centre measured from the interacting galaxies, $m_i$ and $\vect{r}_i$ are the mass and position vector of the $i$-th SPH particle.
This position is then used as centre of a new sphere, within which the mass centre is again calculated, the process is repeated until convergence.

The orbital velocity of the TDG, respectively the simulation box is calculated as mass-weighted average of all stellar and gas SPH-particles within $r_{\smallsub{TDG}}$.
This results in initial positions of the TDGs relative to the interacting host galaxies of $(X,Y,Z)~=~(59.9,-133.9,69.8)~\mathrm{and}~(30.2,289.4,114.1)\,\mathrm{kpc}$ 
and orbital velocities of $\vect{v}_\mathrm{orb}~=~(104.7,-29.6,12.2)$ and $(-35.2,117.8,64.8)\,\mathrm{km\, s}^{-1}$ for the candidates~2 and 3, respectively.

All particles within a box of 22 kpc side length, centred at the mass centre of the TDG, are then selected as the initial data.
To ensure a correct data mapping at the edges of the simulation box, this box is 10\% lager then the later simulation box.

\subsection{Initial gas distribution}\label{sub:inigas}

As the initial conditions for the presented simulations originate from an SPH-simulation, the gas properties are mapped to the AMR grid of \flash~
via the standard kernel function of \textsc{gadget2} \citep{springel05}:

\begin{equation}\label{equ:kernel}
  W(r,h) = \frac{8}{\pi~h^3} \times \begin{cases}
  1-6\left(\frac{r}{h}\right)^2+6\left(\frac{r}{h}\right)^3 	& \dots~ 0 \le \frac{r}{h} \le 0.5 \\
  2 \left( 1-\frac{r}{h}\right)^3 				& \dots~ 0.5 \le \frac{r}{h} \le 1 \\
  0 								& \dots~ \frac{r}{h} > 1,
  \end{cases}
\end{equation}
where $r$ is the distance of a particle to a given point and $h$ is the smoothing length.
Throughout this work $h$ is defined as the distance of the 50th nearest particle to the centre of a given cell, except in cases where more than 50 particles would fall within a grid cell, 
then all particles within this cell are mapped onto it.
For example, the initial density of a grid cell $\rho_\mathrm{cell}$ is then calculated by
\begin{equation}\label{equ:densitymap}
 \rho_\mathrm{cell} = \sum\limits_{i=1}^\mathrm{N} m_{i} W(r_{i},h),
\end{equation}
where $m_i$ is the mass of an SPH gas particle $r_i$ is its distance from the cell centre and $N$ is the number of considered particles.

To account for the pre-enrichment of the ISM the initial metallicity of the TDG and the tidal arm, is set to $Z=0.3~\mathrm{Z}_\odot$ with solar element abundance ratios.

\subsection{Initial stellar population}

Within the presented simulations star clusters are treated as particles, therefore the pre-existing stellar population is initialised with the positions, velocities and masses of the stellar SPH particles.
The stellar particles within the original simulation have a constant mass of $27500~\msol$ each.
Therefore, the 338 stellar particles within the simulation box of candidate~2 provide a total initial stellar mass of $9.295\times10^{6}~\msol$ 
and $2.20\times10^{6}~\msol$ for the 80 stellar particles within the simulation box of candidate~3.

As for the ISM the metallicity of the old stellar population, already formed within the interacting galaxies, is set to $Z=0.3~\mathrm{Z}_\odot$ with solar element abundance ratios, 
because the tidal arms stretch from the outermost stellar disk with the lowest disk abundance.

\subsection{Simulation setup}\label{sub:setup}

The initial gas distribution is mapped, according to the description of Section~\ref{sub:inigas}, to the AMR-grid of the 20~kpc cubed simulation box.
Thereby, we used 5 levels of refinement, resulting in a maximal resolution of 78~pc.

\begin{figure}
 \centerline{\includegraphics[width=\columnwidth]{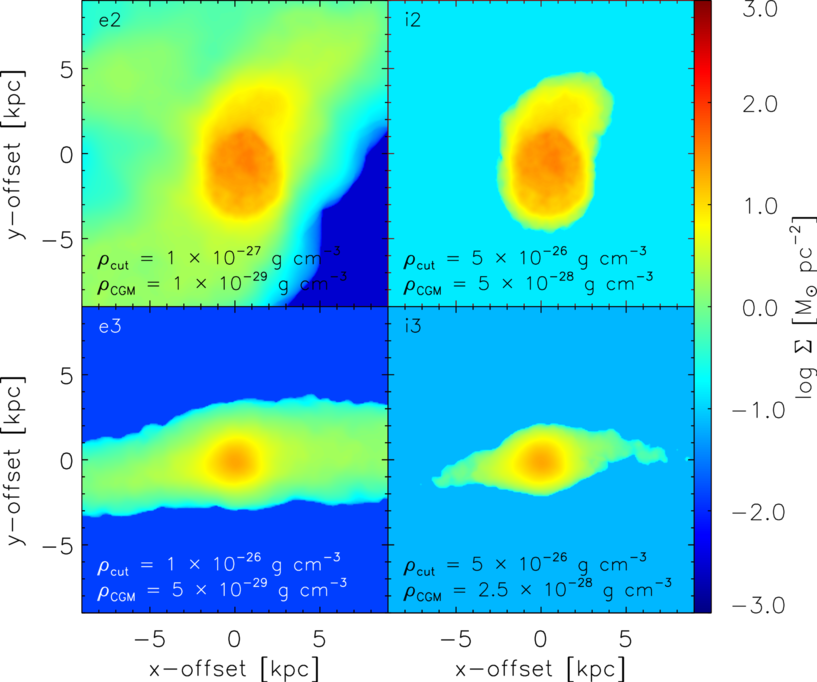}}
 \caption{Initial ($t_\mathrm{sim}=0\,\mathrm{Myr}$) gas surface density, integrated along the z-axis, 
 of the embedded (\textit{e}) and isolated (\textit{i}) models of candidate~2 (top) and candidate~3 (bottom).
 The cutoff density $\rho_\mathrm{cut}$ and CGM density $\rho_\mathrm{CGM}$ are indicated in each panel.}
 \label{fig:surfDens_ini}
\end{figure}

Depending on the considered TDG candidate and the extent of the corresponding section of the tidal arm this process can lead to an almost completely refined simulation box.
In order to avoid a fully refined simulation box, already at the start of the simulation runs, the density distribution is truncated at a lower density $\rho_\mathrm{cut}$.
The resulting gas distribution is then embedded in a homogeneous CGM.
Thereby, the CGM parameters are chosen so that they approximately resemble the Milky Ways hot X-ray halo with $T\approx10^6\,\mathrm{K}$ and $n\approx10^{-4}\,\mathrm{cm}^{-3}$, e.g. \citet{gupta12}.
Keeping the CGM temperature fixed at $10^6\,\mathrm{K}$, the density is adjusted so that the pressure within the tidal arm is slightly higher then the pressure within the CGM.
This prevents additional forces onto the tidal arm which could trigger its collapse.
The cutoff density $\rho_\mathrm{cut}$ and the density of the CGM $\rho_\mathrm{CGM}$ are listed in Table~\ref{tab:runs} for the different simulation runs.
The effect of the different density cuts on the mass distribution can be seen in Figure~\ref{fig:surfDens_ini}.

\begin{table}
 \centering
 \caption[Initial properties of the different models]{Initial properties of the different models and the implied processes. Column 1: Model; Columns 2-4: Included environmental effects;
  Columns 5-6: Density $\rho_\smallsub{CGM}$ of the circumgalactic medium and the cutoff density $\rho_\smallsub{cut}$ of the initial density distribution.}
 \label{tab:runs}
 \begin{tabular}{l|cc|ccc}
 \hline
 \multicolumn{1}{c|}{} 		& \multicolumn{2}{c|}{CGM}			& \multicolumn{3}{c}{Environment}	\\
 \multicolumn{1}{c|}{model}	& $\rho_\smallsub{CGM}$ & $\rho_\smallsub{cut}$	& tidal & tidal	& ram			\\
 \multicolumn{1}{c|}{} 		& [g cm$^{-3}$] 	& [g cm$^{-3}$]		& arm 	& field	& pressure		\\
 \hline
 e2	& $1\times10^{-29}$	& $1\times10^{-27}$	& \checkmark 	& ---		& ---		\\
 e2t	& $1\times10^{-29}$	& $1\times10^{-27}$	& \checkmark 	& \checkmark 	& ---		\\
 e2tr	& $1\times10^{-29}$	& $1\times10^{-27}$	& \checkmark 	& \checkmark	& \checkmark	\\
 i2	& $5\times10^{-28}$	& $5\times10^{-26}$	& ---		& ---		& ---		\\
 i2t	& $5\times10^{-28}$	& $5\times10^{-26}$	& ---		& \checkmark	& ---	 	\\
 i2tr	& $5\times10^{-28}$	& $5\times10^{-26}$	& ---		& \checkmark 	& \checkmark 	\\
 \hline
 e3	& $5\times10^{-29}$	& $1\times10^{-26}$	& \checkmark 	& ---		& ---		\\
 e3t	& $5\times10^{-29}$	& $1\times10^{-26}$	& \checkmark 	& \checkmark 	& ---		\\
 e3tr	& $5\times10^{-29}$	& $1\times10^{-26}$	& \checkmark 	& \checkmark	& \checkmark	\\
 i3	& $2.5\times10^{-28}$	& $5\times10^{-26}$	& ---		& ---		& ---		\\
 i3t	& $2.5\times10^{-28}$	& $5\times10^{-26}$	& ---		& \checkmark	& ---	 	\\
 i3tr	& $2.5\times10^{-28}$	& $5\times10^{-26}$	& ---		& \checkmark 	& \checkmark 	\\
 \hline
 \end{tabular}
\end{table}

As boundary conditions \flash's standard outflow boundaries are used.
These boundary conditions are insensitive to the flow direction across the boundary, therefore they also allow for a flow into the simulation box.
This inflow is used to mimic the flow along the tidal arm across the physical boundaries of the simulation box and, by this, 
for the first time the TDG evolution within the tidal arm is self-consistently coupled to the gas and stellar dynamics of the tidal-arm environment.

In a series of six simulation runs for each TDG candidate we expose the TDGs to different combinations of environmental effects to study their impact on the formation and evolution of TDGs.
In three of these runs the TDG is embedded in the tidal arm, indicated by model names starting with \textit{e}, 
the remaining three runs are consider to be isolated from the tidal arm, denoted by \textit{i}.
Throughout this work the term isolated refers to the non-existence of the tidal arm rather than the spatial isolation from any neighbouring galaxy.
The inclusion of the different environmental effects is indicated by the suffix of the model name, where \textit{t} indicates the presence of the tidal field and \textit{r} the activity of ram pressure.
Among the models of candidate~2, e2tr is the most realistic model of a young TDG, as it includes the tidal arm and the effects of the tidal field and ram pressure.
Contrary to that, the model i2 can be considered as a model of a DM free and fully isolated DG, without an additional gas reservoir or any environmental influences.
An overview of the different models and some of their initial parameters and settings is provided in Table~\ref{tab:runs}. 
The discrepancies in the CGM density between the e and i models is caused by the necessity to keep the isolated DGs with the same central density and mass bound by the external gas pressure. 
This is not necessary for the embedded models due to the tidal-arm gas dynamics.

\section{Results}\label{sec:results}		

\subsection{Mass assembly}\label{sub:massas}

The presented simulations start $1.5\,\mathrm{Gyr}$ after the first close encounter of the parent galaxies, 
at this time already a Jeans unstable gas cloud embedded in the tidal arm has formed (see Figure~\ref{fig:surfDens_ini}).
During the first few tens of Myr a high density core is developing within the cloudy distribution of gas, resulting in a short phase of low SF.
As the core accumulates more and more gas, that condenses dissipatively, it forces the surrounding matter to collapse. 
During this phase arm-like structures are created which effectively funnel the gas to the centre of the TDG.
This leads to very high central densities, a high SFR, and a very short gas consumption time. 

After the collapse-like mass assembly, which ends at a simulation time of $t_\mathrm{sim}=300\,\mathrm{Myr}$, the majority of gas is converted into stars.
Further mass accumulation during the rest of the simulation time is caused by accretion along the tidal arm, if it is taken into account in the simulation run (\textit{e} models).
At this time the isolated models are reaching their total final mass.

\begin{figure*}
 \centerline{\includegraphics[width=\columnwidth]{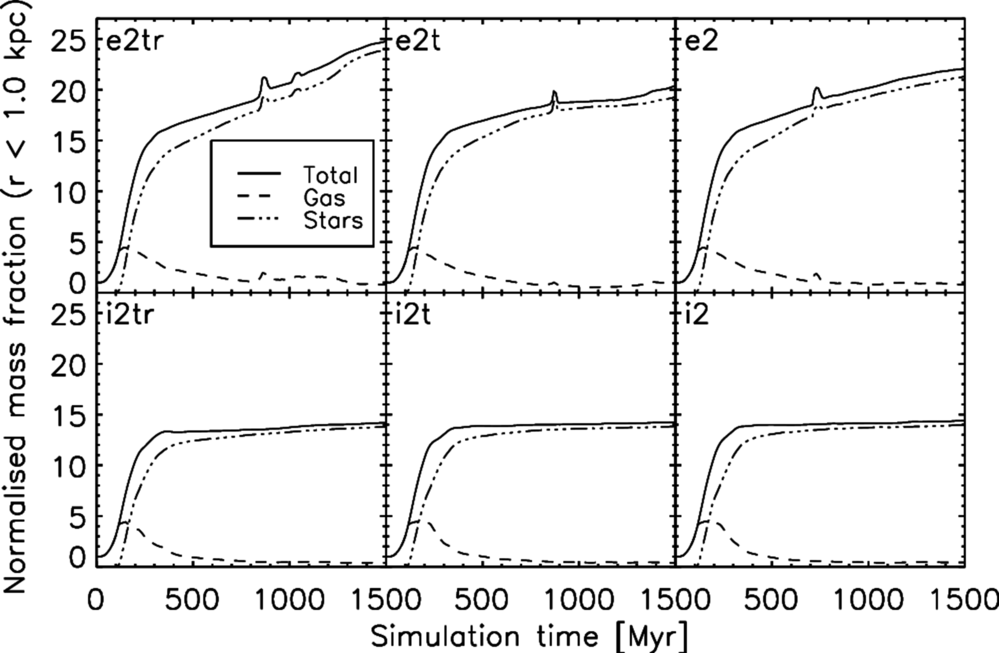} \hfill
 \includegraphics[width=\columnwidth]{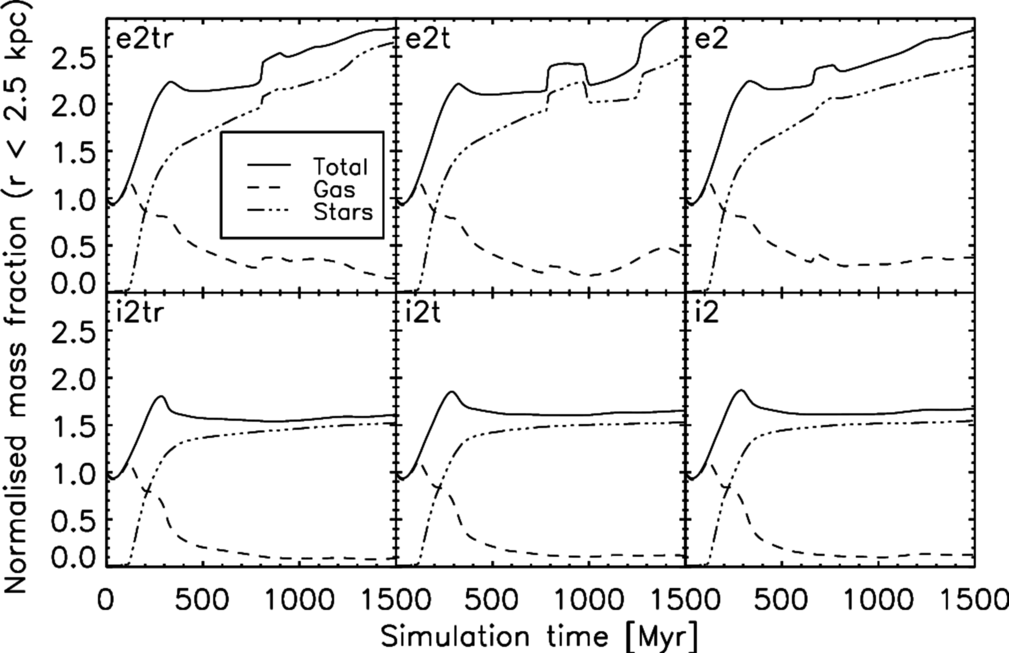}}
 \caption{Relative mass fractions, normalised to the total initial mass, within 1.0 kpc (left) and 2.5 kpc (right) from the centre of mass for the different models of candidate~2. 
 In the top row the models embedded in the tidal arm are shown and the corresponding isolated models are in the bottom row.
 The line styles in all panels are identical and indicated in the top left panels. Note the different scaling on the y-axis.}
 \label{fig:mfrac25}
\end{figure*}

\begin{figure*}
 \centerline{\includegraphics[width=\columnwidth]{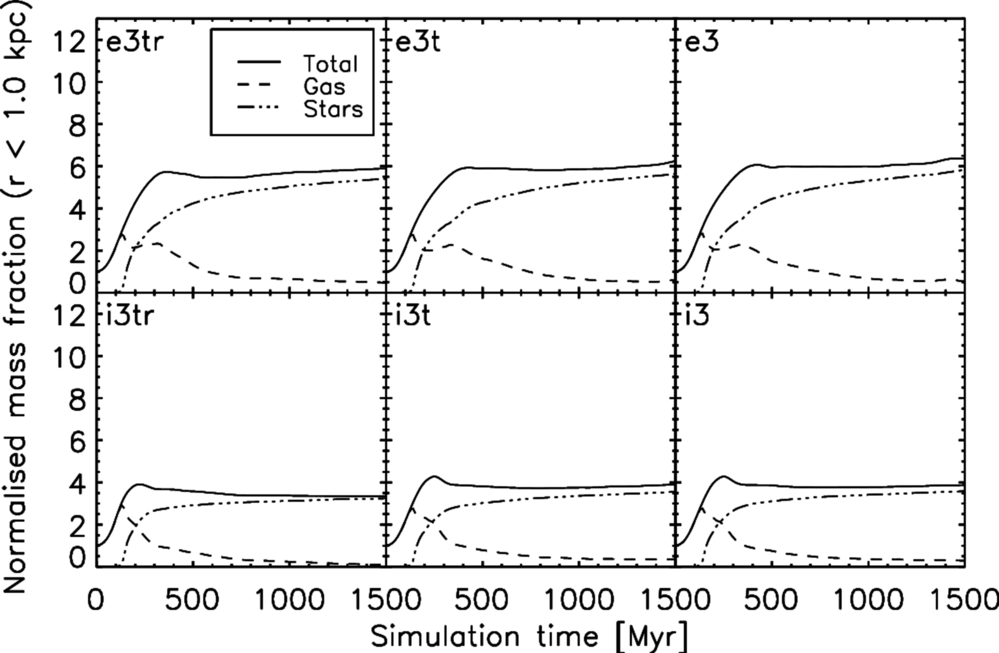} \hfill
 \includegraphics[width=\columnwidth]{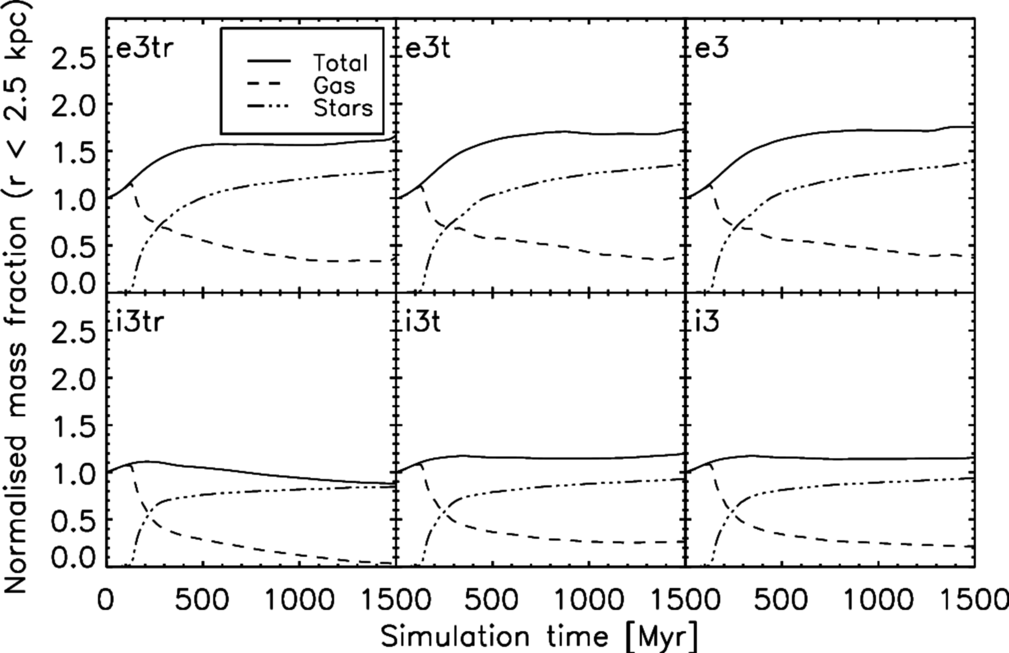}}
 \caption{Same as Figure~\ref{fig:mfrac25} for the models of candidate~3. }
 \label{fig:mfrac25_s3}
\end{figure*}

Figures~\ref{fig:mfrac25} and \ref{fig:mfrac25_s3} show the evolution of the relative mass fractions, normalised to the total initial mass, 
within spheres of 1.0 (left panels) and 2.5 kpc (right panels) radius around the TDGs centre for the candidate~2 and 3, respectively.
During the collapse the total mass within 2.5 kpc increases to approximately $1.8-2.3$ times the initial mass for the candidate~2 and $1.1-1.5$ times for candidate~3, 
depending on the presence of the tidal arm. 
During the initial collapse the central regions ($r\le 1.0 \,\mathrm{kpc}$) are growing by factors of $\sim18$ for the embedded and $\sim15$ for the isolated models of candidate~2.
Due to the less extended tidal arm surrounding candidate~3, these models only increase their central mass by a factor of $4-6$.
The transition from a gas to a stellar mass dominated system is indicated by the intersection of the dashed (gas) and dot-dashed (stars) lines at $t_\mathrm{sim}\approx150\,\mathrm{Myr}$.
The influence of the tidal arm for the mass assembly can clearly be seen as the embedded models continue to increase their mass until the end of the simulation time.
The embedded models of candidate~2 increase their mass within the inner 1.0 kpc of radius by $20-25$ times the initial mass.

\begin{figure*}
 \centerline{\includegraphics[width=\columnwidth]{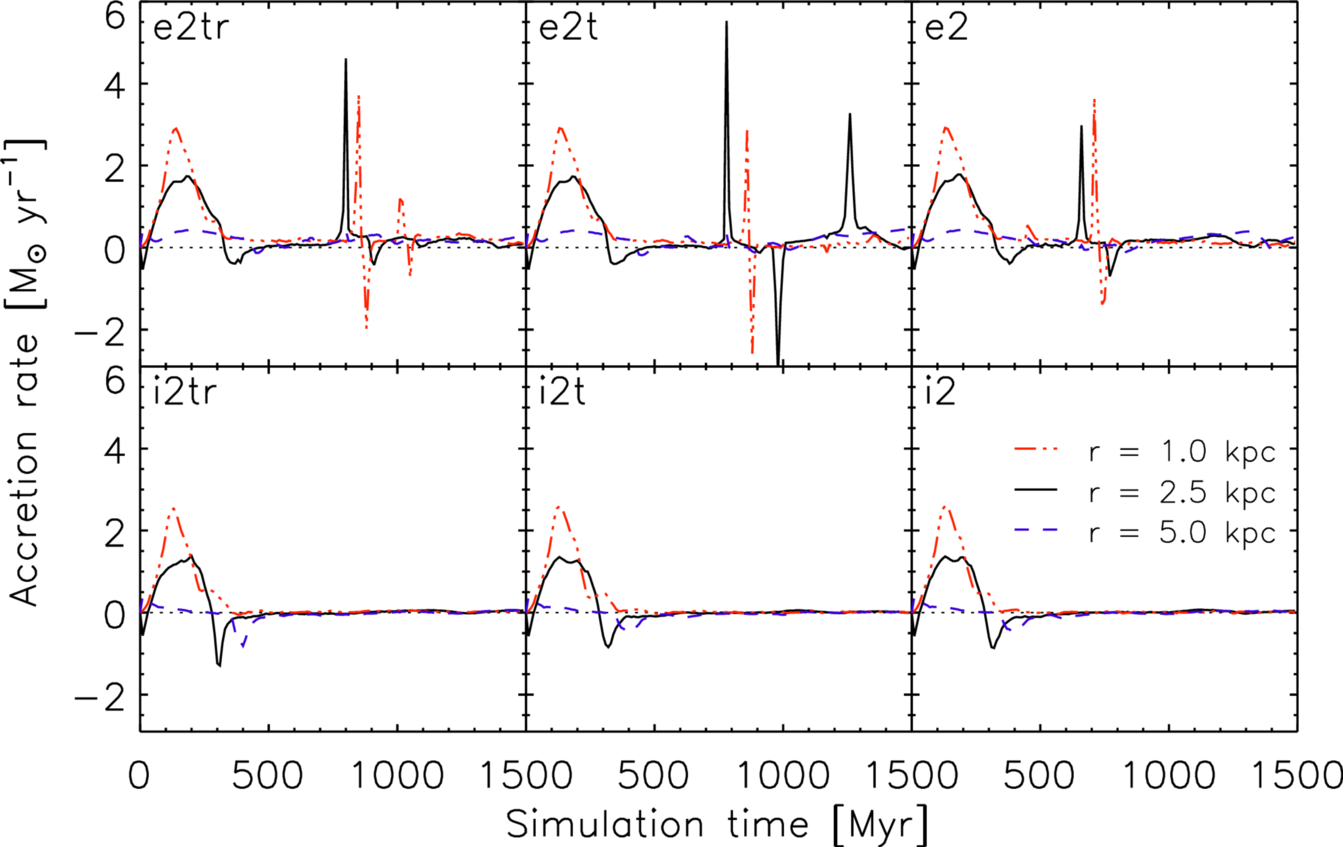} \hfill 
 \includegraphics[width=\columnwidth]{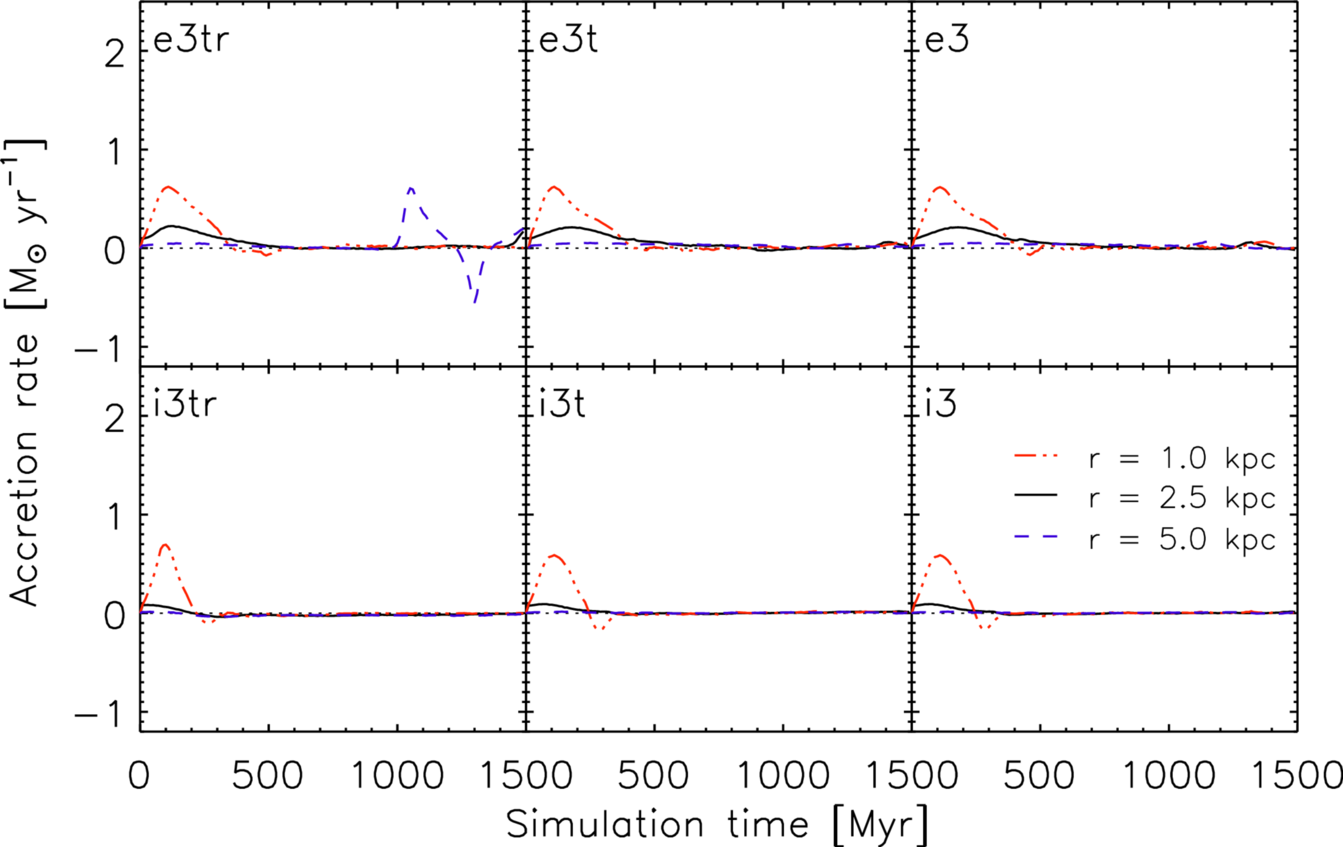}}
 \caption{Accretion rates through spheres with $r~=~1.0,~2.5~\mathrm{and}~5.0~\mathrm{kpc}$ around the mass centre of the TDGs.
 The left panels correspond to the models of candidate~2 and the models of candidate~3 are shown in the right panels. The line styles are indicated in the bottom right panels.
 }
 \label{fig:accretion}
\end{figure*}

The accretion rates in $\usfr$, averaged over 10~Myrs, through spherical surfaces with radii of 1.0, 2.5 and 5.0~kpc around the TDG are shown in Figures~\ref{fig:accretion}.
The accretion rate through the 5~kpc sphere generally remain low, below $1.0~\usfr$ in all models of candidate~2.
During the initial collapse the embedded models of this candidate are reaching accretion rates of $\sim3\,\mathrm{and}\,\sim2~\usfr$ at distances of 1.0 and 2.5~kpc, respectively.
As the tidal arm is less extended in the embedded models of candidate~3 the accretion rates are much smaller compared to the models of candidate~2.
Through the outer 5~kpc sphere maximal accretion rates of $\sim0.1~\usfr$ are reached and for the 2.5 and 1~kpc spheres accretion rates of $\sim0.7$ and $\sim0.2~\usfr$ are reached during the collapse phase.

The isolated models behave qualitatively similar to the embedded models during this phase, with slightly lower accretion rates.
For the central sphere their accretion rate drops to zero after roughly $300-500$~Myr, as, due to the absence of the tidal arm, no further material is available for accretion.
The embedded models continue to grow in mass by the accumulation of matter along the tidal arm with accretion rates around $0.5~\usfr$ and $0.05~\usfr$ for the candidates~2 and 3, respectively.

\begin{figure}
 \centerline{\includegraphics[width=\columnwidth]{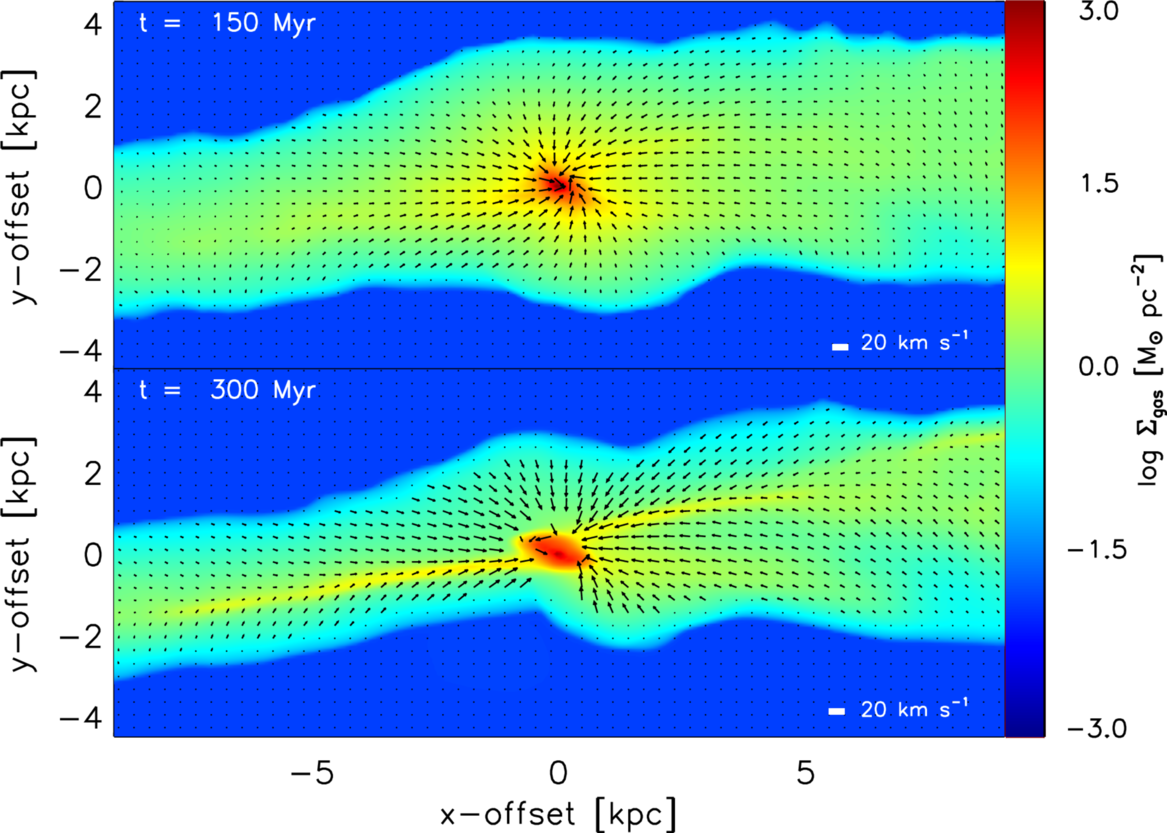}}
 \caption{Density weighted velocity field  of the tidal arm integrated along the line of sight of model e3t at $t_\mathrm{sim}=150~\mathrm{and}~300\,\mathrm{Myr}$.
 For clarity velocities greater than $20\,\mathrm{km\,s}^{-1}$ are not displayed. The white line in the bottom right corner indicates a velocity of $20\,\mathrm{km\,s}^{-1}$}
 \label{fig:velfield}
\end{figure}

The dynamics within the tidal arm of the model e3t at $t_\mathrm{sim}=150~\mathrm{and}~300~\mathrm{Myr}$ is illustrated in Figure~\ref{fig:velfield}.
At $t_\mathrm{sim}=150~\mathrm{Myr}$, the peak of the collapse phase, an almost spherical accretion is present, 
while at $t_\mathrm{sim}=300~\mathrm{Myr}$, close to the end of the collapse phase, streams with higher densities are formed within the tidal arm along which the gas is transported towards the TDG.

The thin spikes in Figure~\ref{fig:accretion} correspond to captured respectively ejected substructures like dense gas clouds and/or massive star clusters, 
which are formed in the close vicinity of the TDG, at distances beyond $2.5\,\mathrm{kpc}$ form their centre.
These spikes occure exclusively in candidate 2 that is much more affected by gas streams and mass assembly from the tidal arm than candidate 3. 
That these spikes are more pronounced at the 2.5 kpc radius from the mass center but less at 5 kpc distance means that clumps fall towards the innermost region. 
And that the clumps form between these radii can be concluded from the total mass increase (see Figure~\ref{fig:mfrac25}).

The mass distribution among gas and stars is listed in Table~\ref{tab:results} for the final simulation time ($t_\mathrm{sim}=1500\,\mathrm{Myr}$).

\begin{table*}
 \centering
 \caption[Model properties at the final simulation time]{Model properties at the final simulation time ($t_\mathrm{sim}=1\,500~\mathrm{Myr}$). Column 1: Model; 
  Columns 2-4: Gas mass $M_\mathrm{gas}$, total mass of newly formed and pre-existing stars $M_\mathrm{star}$ and gas fraction $f_\mathrm{g}=M_\mathrm{gas}/\left(M_\mathrm{gas}+M_\mathrm{star}\right)$ 
  within $r=2.5\mathrm{~kpc}$ from the centre of the TDG;
  Columns 5-7: $\mathrm{SFR}$ at the final simulation time, the peak SFR during the collapse and average SFR over the whole simulation time.}
 \label{tab:results}
 \begin{tabular}{l|cccccc}
 \hline
 \multicolumn{1}{c|}{model}	& $M_\mathrm{gas}$	& $M_\mathrm{star}$	& $f_\mathrm{gas}$	& $\mathrm{SFR}$ & $\mathrm{SFR_{peak}}$	& $<\mathrm{SFR}>$	\\
 \multicolumn{1}{c|}{} 		& [$10^8$ M$_\odot$]	& [$10^8$ M$_\odot$]	& 			& 		\multicolumn{3}{c}{[$\usfr$]}		\\
 \hline
 e2	& 0.98 	& 7.32	& 0.12	& 0.18	& 3.46	& 0.60	\\
 e2t	& 1.04	& 6.60	& 0.14	& 0.35	& 3.55	& 0.65	\\
 e2tr	& 0.39	& 7.00	& 0.05	& 0.21	& 3.57	& 0.68	\\
 i2	& 0.34	& 4.07	& 0.08	& 0.07	& 2.95	& 0.39	\\
 i2t	& 0.31	& 4.03	& 0.07	& 0.05	& 3.04	& 0.39	\\
 i2tr	& 0.23	& 4.01	& 0.05	& 0.06	& 2.96	& 0.40	\\
 \hline
 e3	& 0.41 	& 1.56	& 0.21	& 0.05	& 1.22	& 0.15	\\
 e3t	& 0.42	& 1.54	& 0.21	& 0.05	& 1.23	& 0.15	\\
 e3tr	& 0.42	& 1.45	& 0.22	& 0.04	& 1.27	& 0.14	\\
 i3	& 0.22	& 0.96	& 0.19	& 0.02	& 1.06	& 0.09	\\
 i3t	& 0.27	& 0.95	& 0.22	& 0.02	& 1.09	& 0.09	\\
 i3tr	& 0.04	& 0.87	& 0.04	& 0.01	& 1.30	& 0.08	\\
 \hline
 \end{tabular}
\end{table*}

\subsection{Star formation}

Regardless of the environmental impact factors, i.e. the tidal field and ram pressure or the presence of the embedding tidal arm, the star-formation history (SFH) of all runs is similar and can be described by three different stages. 
First, an initial SF episode followed by a strong starburst triggered by the collapse of the proto-TDG and a subsequent long-living phase of self-regulated SF.
These different SF episodes follow the evolutionary phases as described in the previous Section \ref{sub:massas}.

The low initial SFR is caused by the early built-up of the TDGs core. During this phase the SFR remains low at a level of about $10^{-2}~\usfr$.

As the evolution of the TDG proceeds, the surrounding material collapses within a free-fall time, 
triggering the starburst at $t_\mathrm{sim}\approx130\,\mathrm{Myr}$ for a period of about 100 Myr before the SFR starts to decline again.
Thereby, SFRs of up to $3.5~\mathrm{and}~1.2~\usfr$ are reached for the candidate~2 and 3, respectively, 

Depending on the availability of gas, the SFR approaches an approximate equilibrium level after 300 to 500 Myr. 
In this phase, the SFR of the models embedded in the tidal arm stays around $0.5~\mathrm{and}~0.1~\usfr$ for candidate~2 and 3, respectively,
whereas, the self-regulated equilibrium SFR of the isolated models, lacking further material to be accreted from the tidal arm reservoir, is roughly one half to one order of magnitude lower.

\begin{figure*}
 \centerline{\includegraphics[width=\columnwidth]{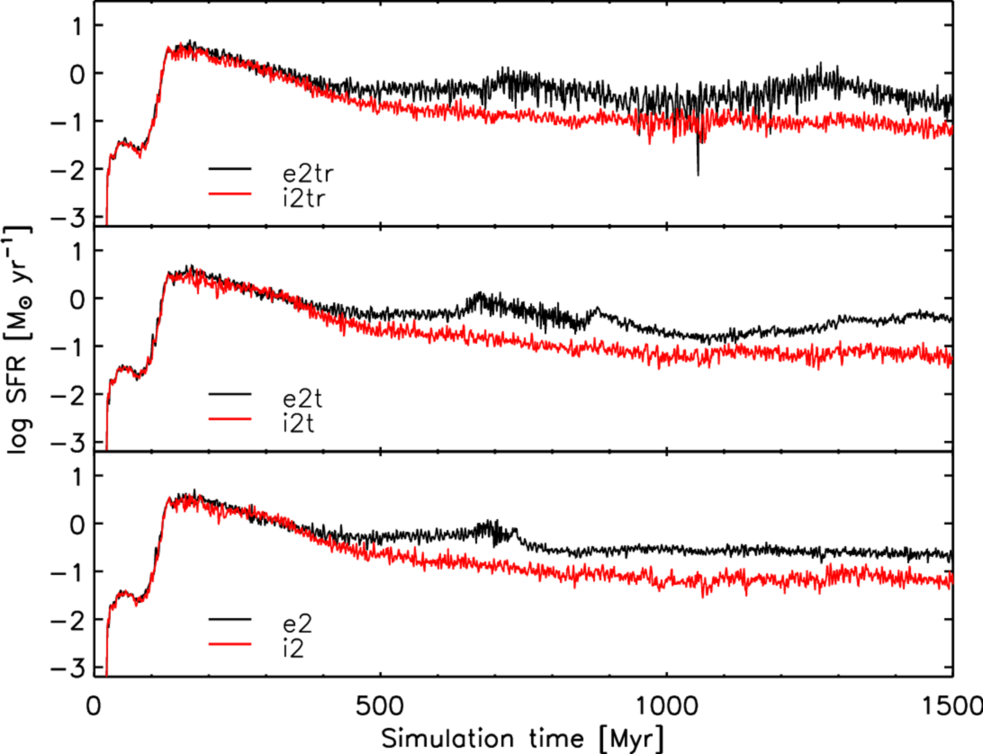}  \hfill 
 \includegraphics[width=\columnwidth]{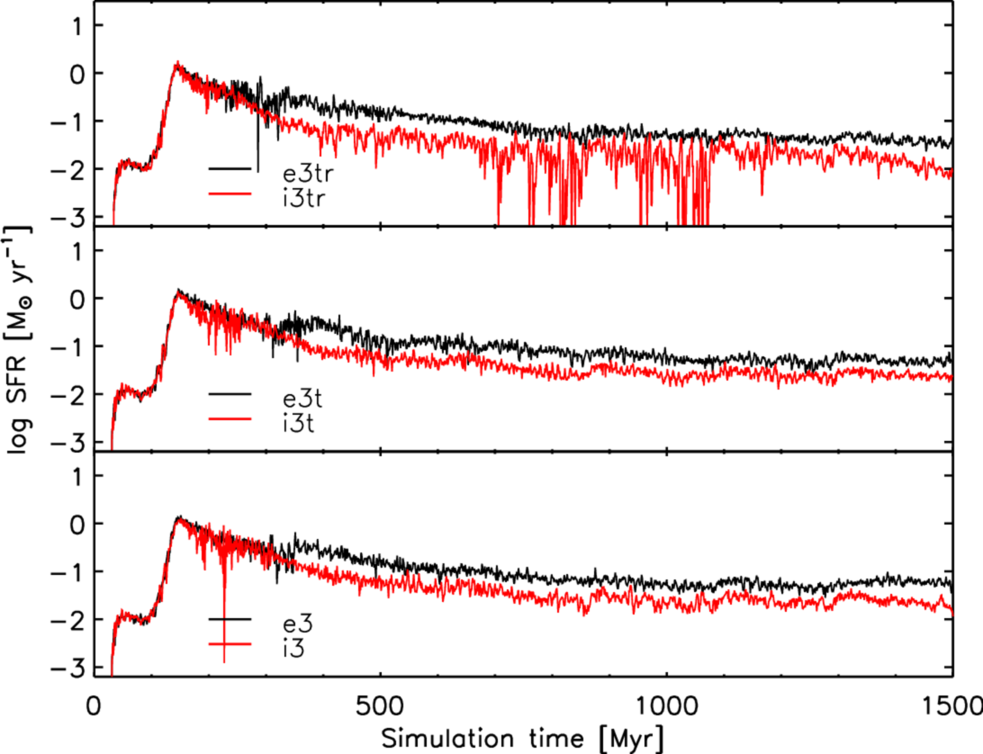}}
 \caption{Comparison of the SFRs of the embedded and isolated models of candidate~2 (left) and candidate~3 (right). 
 From top to bottom the models are sorted according to the included environmental effects.
 The top panel includes the tidal field and ram pressure, the middle panel only includes the tidal field and the bottom panel none of these effects.
 }
 \label{fig:sfr}
\end{figure*}

In Figure~\ref{fig:sfr} the SFHs of the different simulation runs are compared, sorted by the included environmental effects, 
from top to bottom, including the tidal field and ram pressure, the tidal field only and none of these effects.
The oscillation in the SFR or the model i3tr between $700$ and $1100~\mathrm{Myr}$ is caused by the removal of gas due to RPS.
The in-situ formed stars dominate the stellar mass of both TDG candidates, until the end of the simulation at $t_\mathrm{sim}~=~1500~\mathrm{Myr}$, 
the mass fraction of the pre-existing stellar population decreases to $1\,\%$ of the total stellar mass for the embedded models and to $2\,\%$ for the isolated models.
For comparison the final SFRs, peak SFRs and average SFRs are compiled in Table~\ref{tab:results}. 

\subsection{Environmental influence}

Ram pressure caused by the motion of the TDG thought the CGM of the parent galaxies scales with the square of its orbital velocity 
and the strength of the tidal field is weakening with the cube of the distance to these galaxies.
Therefore, the environmental effects are strongly affected by the actual orbit of the TDG candidate~2 around its interacting host galaxies.
With a distance of $162.5~\mathrm{kpc}$ from the mass centre of the interacting galaxies and an orbital velocity of $109.5~\mathrm{km\,s}^{-1}$, at the initial simulation time, 
the TDG candidate~2 remains on a bound eccentric orbit around its parents, one can calculate its orbital period to amount to $8.9~\mathrm{Gyr}$.
Until the end of the simulation time of $1500~\mathrm{Myr}$ this TDG is reaching a distance of $223.2~\mathrm{kpc}$ 
at a velocity of $71.4~\mathrm{km\,s}^{-1}$ as it is approaching its orbital apocentre.

The TDG candidate~3, at an initial distance of $312.5~\mathrm{kpc}$ with an orbital velocity of $187.0~\mathrm{km\,s}^{-1}$ is on an unbound orbit, escaping from the parent galaxies.
At the final simulation time it reaches a distance of $575.1~\mathrm{kpc}$ with a velocity of $167.9~\mathrm{km\,s}^{-1}$.

\subsubsection{The tidal field}

The compressive respectively disrupting effect of tidal forces exerted on TDGs by their parent galaxies' gravitational potential are believed 
to play a crucial role in the formation and evolution of TDGs \citep{duc04}.
This effect is investigated by comparing the simulation runs e2 to e2t and i2 to i2t.
The models e2 and i2 lack of any tidal accelerations and ram pressure, whereas the models e2t and i2t are exposed to the tidal field.
The embedded models of candidate 2 (e2 and e2t), do not show strong differences in their SFRs 
(Figure~\ref{fig:sfr}) or gas and stellar mass fractions (Figure~\ref{fig:mfrac25})
which could be attributed to the tidal field.

\begin{figure}
  \centerline{\includegraphics[width=\columnwidth]{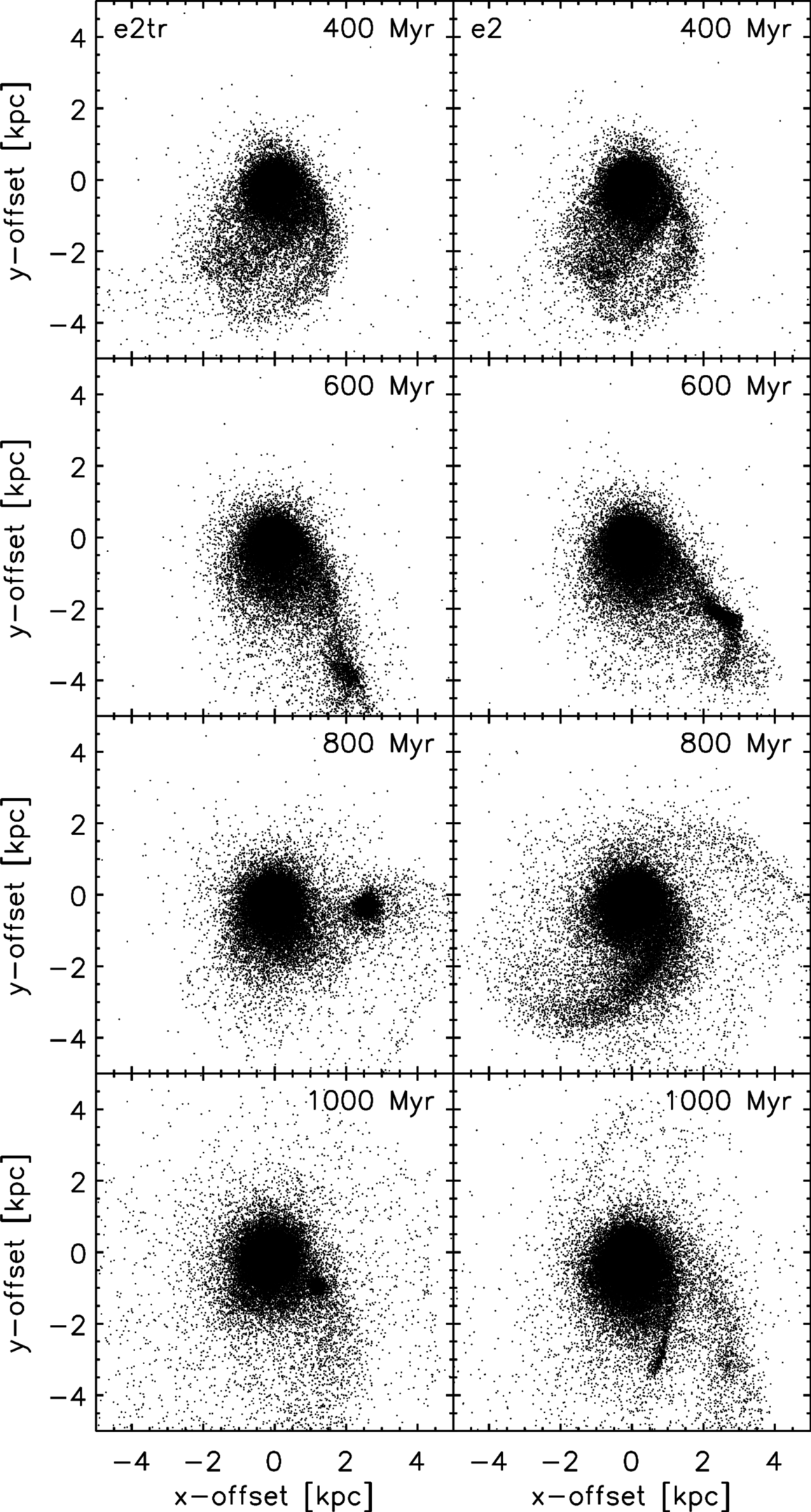} }
  \caption{Face-on stellar distribution of the models e2tr (left) and e2 (right) in the simulation time interval $400 - 1000~\mathrm{Myr}$ in steps of $200~\mathrm{Myr}$ (from top to bottom).
	    Each panel is centred on the mass centre of the TDG candidate.}
 \label{fig:c2stars}
\end{figure}

The differences in the two embedded models of candidate~2 emerging after $600-700\,\mathrm{Myr}$ 
are attributed to the different accretion histories of these models.
A super star clusters with a mass in the order $10^7 \msol$ forms close to the TDG within the tidal arm at 
different strengths, i.e. density and streaming motion, and times and is therefore accreted at different times, 
corresponding to jumps in the right hand panel of Figure~\ref{fig:mfrac25} and the thin spikes in the left panel of Figure~\ref{fig:accretion}.
The resulting structural differences in the stellar distribution are illustrated in Figure \ref{fig:c2stars} for the models e2tr and e2 in the simulation time interval $400-1000~\mathrm{Myr}$.

The two isolated models i2 and i2t do not show any significant differences in their evolution, neither in their SFR nor in their mass composition.

With $f_\mathrm{gas}=0.21$ and $\mathrm{SFR}=0.05$ after 1.5~Gyrs for both corresponding embedded models of candidate~3 (e3 and e3t) 
show an even smaller impact of the parent galaxies' tidal field on their evolution.
Due to the unbound orbit of candidate~3 the gravitational influence of the parent galaxies decreases with time.

\subsubsection{Ram pressure}

Ram pressure stripping (RPS) is thought to be one of the main drivers of DG transformation, from gas rich to gas poor, even for DM dominated DGs.
Therefore, TDGs, which are supposed to be free of DM, should be especially sensitive to RPS.
This is only partly true, as young TDGs in the process of formation and during their early evolution, as those simulated in this work, are embedded in a gaseous tidal arm.

Before the ram pressure can directly act on the TDG itself the tidal arm has to be dissolved and mostly decouple from the TDG.
This is causing a significant delay time before a young TDG is actually affected.
During this delay time, typically a few 100~Myr within the presented simulations, a large fraction of gas is already converted into stars, 
leaving behind a stellar dominated system even before RPS of the TDG starts.

As long as the arm and the TDG are not completely separated, some of the TDGs gas is continuously stripped while simultaneously the TDG is refueled by material accreted along the tidal arm.
Due to the compression of the tidal arm by ram pressure the accretion of gas along the tidal arm can be higher in these models compared to those without ram pressure.
Thereby, the accretion rate can be even higher than the mass loss rate due to RPS.

After the majority of low-density gas has been removed from the upwind side of the TDG, ram pressure starts to influence the high-density disk.
Since the TDG's spin vector should be oriented perpendicular to the orbital plane of the tidal arm, ram pressure is typically acting edge-on onto the TDG's 
gas disk leading to its deformation and compression while some of its gas is lost.

\begin{figure}
 \centerline{\includegraphics[width=\columnwidth]{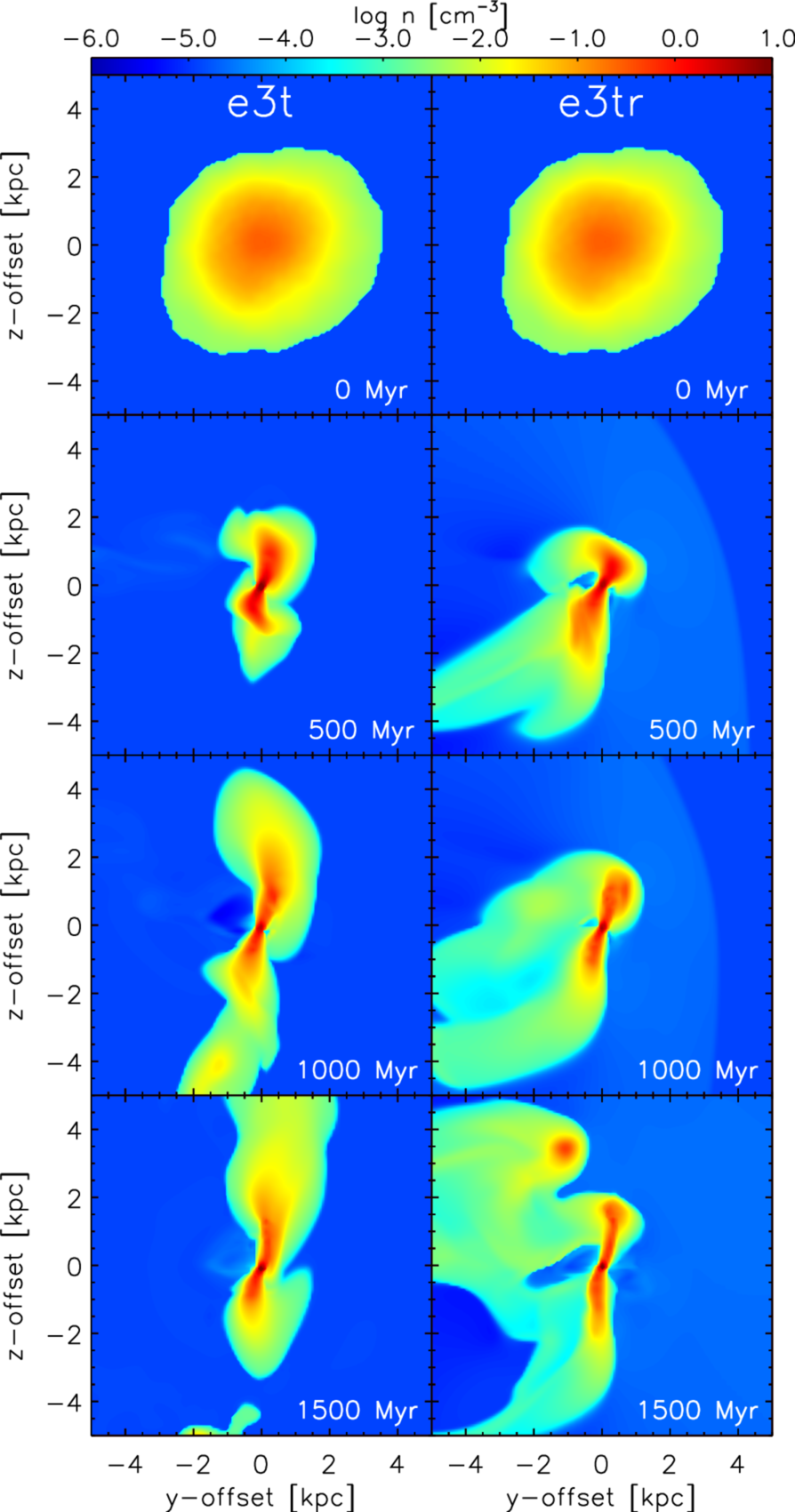}}
 \caption{Density slices thought the centre of mass of the TDG for model e3t (left) and e3tr (right) at $t_\mathrm{sim}~=~0,~500,~1000~\mathrm{and}~1500\,\mathrm{Myr}$ from top to bottom.
 Colour coded is the particle number density in [$\mathrm{atoms~cm}^{-3}$]}
 \label{fig:dens_cut}
\end{figure}

By comparison of the density slices of the two embedded models e3t (left) and e3tr (right) of Figure~\ref{fig:dens_cut} the effect of ram pressure is exemplary illustrated.
Model e3t only feels the tidal field of the parent galaxies, whereas e3tr is additionally exposed to ram pressure.
During the early evolutionary phases ($t_\mathrm{sim}\la500\,\mathrm{Myr}$) the rarefied gas of the tidal arm is pushed around the TDG from the upwind side and stripped away at the lee side.
The high-density TDG gas remains mostly unaffected until most of the low-density gas is removed. 
At this stage ram pressure starts acting on the high density disk of the TDG.
Due to the inclined orientation of the ram pressure a hammerhead-shaped deformation of the gaseous disk appears 
($t_\mathrm{sim}=500\,\mathrm{Myr}$ panel of model e3tr).
Further mass loss due to continuous stripping occurs throughout the whole simulation time at very low rates and is refueled by accretion along the tidal arm.
The small differences in $M_\mathrm{gas}$, between $500\le t_\mathrm{sim}/\mathrm{Myr}\le1500$, 
of the models e3t and e3tr are balanced until the end of the simulation by slightly higher accretion rates along the compressed tidal arm.

As a result of ram pressure the TDGs high density disk in the e3tr model is thinner and elongated towards the negative z-direction compared to the model e3t at $t_\mathrm{sim}=1500\,\mathrm{Myr}$.
The density peak at $z\approx4\,\mathrm{kpc}$ in the $t_\mathrm{sim}~=~1500\,\mathrm{Myr}$ panel of model e3tr does not belong to the TDG itself but is part of the tidal arm.
Due to ram pressure the tidal arm in this model is compressed and bent behind TDG.

The differences in the evolution, both in SFR and mass distribution, of the models e2t and e2tr are attributed to the different accretion histories of these models, 
which is partly influenced by the compression of the tidal arm due to ram pressure.

The only model severely influenced by ram pressure is the model i3tr, a model mostly comparable to a DM-free satellite DG feeling the tidal field of the host galaxy and ram pressure.
This model looses a large fraction of its ISM. Until the end of the simulation time it only retains $4\times 10^5~\msol$ of gas, corresponding to a gas fraction of only 4\%, 
whereas the other isolated models of candidate~3 have gas masses of $\sim2\times 10^6~\msol$ and a corresponding gas fraction of $\sim20\%$.
Due to the lower relative velocity of the TDG with respect to the CGM in the models of the candidate~2, the impact of RPS is less pronounced then in the models of candidate~3.
The model i2tr retains a gas mass of $2.3\times10^{7}\,\msol$ corresponding to a gas fraction of 5\% whereas the gas mass of i2t is $3.1\times10^{7}\,\msol$,
respectively the gas fraction is 7\% and is only marginally larger.


\subsubsection{The tidal arm}

Within the presented simulations, the presence of the additional gas reservoir within the tidal arm shows the strongest impact on the evolution of forming TDGs.
Thereby, it influences the early evolution of TDGs in different ways.
Due to the additional supply of gas, it increases the fraction of gas contained in a TDG and thus also enhances the SFR until the end of the simulation time.
Additionally, it shields the TDG from its parents' CGM, through which it is travelling and thus it is protecting the TDG from the direct influence of ram pressure.
Within the embedded models it appears that the combination of the tidal arm and ram pressure can even increase the gas mass within a TDG.
As the tidal arm gets compressed by ram pressure, the density within the arm is locally increased, allowing for higher accretion rates along the arm.
These effects have not been taken into account in previous numerical studies of TDGs, like those by \citet{smith13,yang14,ploe14,ploe15}.

Initially all models of one candidate start with the same central gas distribution, deviating at densities below $5\times10^{-26}~\cgsdens$,
i.e. the cutoff density of the isolated models (see Table \ref{tab:runs}), resulting in almost equal gas masses within $r=2.5\,\mathrm{kpc}$.

During the initial collapse the mass within that radius remains approximately equal between the embedded and isolated models.
After the collapse of the proto-cloud, around $t_\mathrm{sim}\approx 250~\mathrm{and}~180\,\mathrm{Myr}$, for candidate~2 and 3 respectively,
the gas mass of the isolated models declines significantly stronger compared to the models embedded in the tidal arm.
This behaviour is caused by the lack of surrounding material which could be accreted by the TDGs. 

At the final simulation time ($t_\mathrm{sim}=1500\,\mathrm{Myr}$) the embedded models retained twice as much gas than those in isolation within $r=1.0\,\mathrm{kpc}$.
Going to larger distances the deviations can become as large as a factor of 3 or 5 for $r=2.5\,\mathrm{kpc}$ and $r=5.0\,\mathrm{kpc}$, respectively.

In the embedded and the isolated models, the SFR follows the available amount of gas,
therefore a strong starburst associated with the collapse of the TDG is present.
After the collapse the SFR of the isolated models declines due to gas consumption and the missing gas reservoir, called starvation \citep{larson80}.
As the embedded models are constantly fed with new material along the tidal arm, their SFRs stay at higher levels throughout the remaining simulation time.

This results in a three times higher SFR at the end of the simulations if the tidal arm is included, 
the isolated models of candidate~2 have final SFRs of about $6\times10^{-2} ~\usfr$ compared to $2\times10^{-1} ~\usfr$ if the tidal arm is included. 
The results for candidate~3 a similar, where the embedded models, with SFRs in the order of $5\times10^{-2} ~\usfr$, have 2.5 times higher SFRs at the end of the simulation runs 
compared to $2\times10^{-2} ~\usfr$ for the isolated models.

\begin{figure*}
 \centerline{\includegraphics[width=\columnwidth]{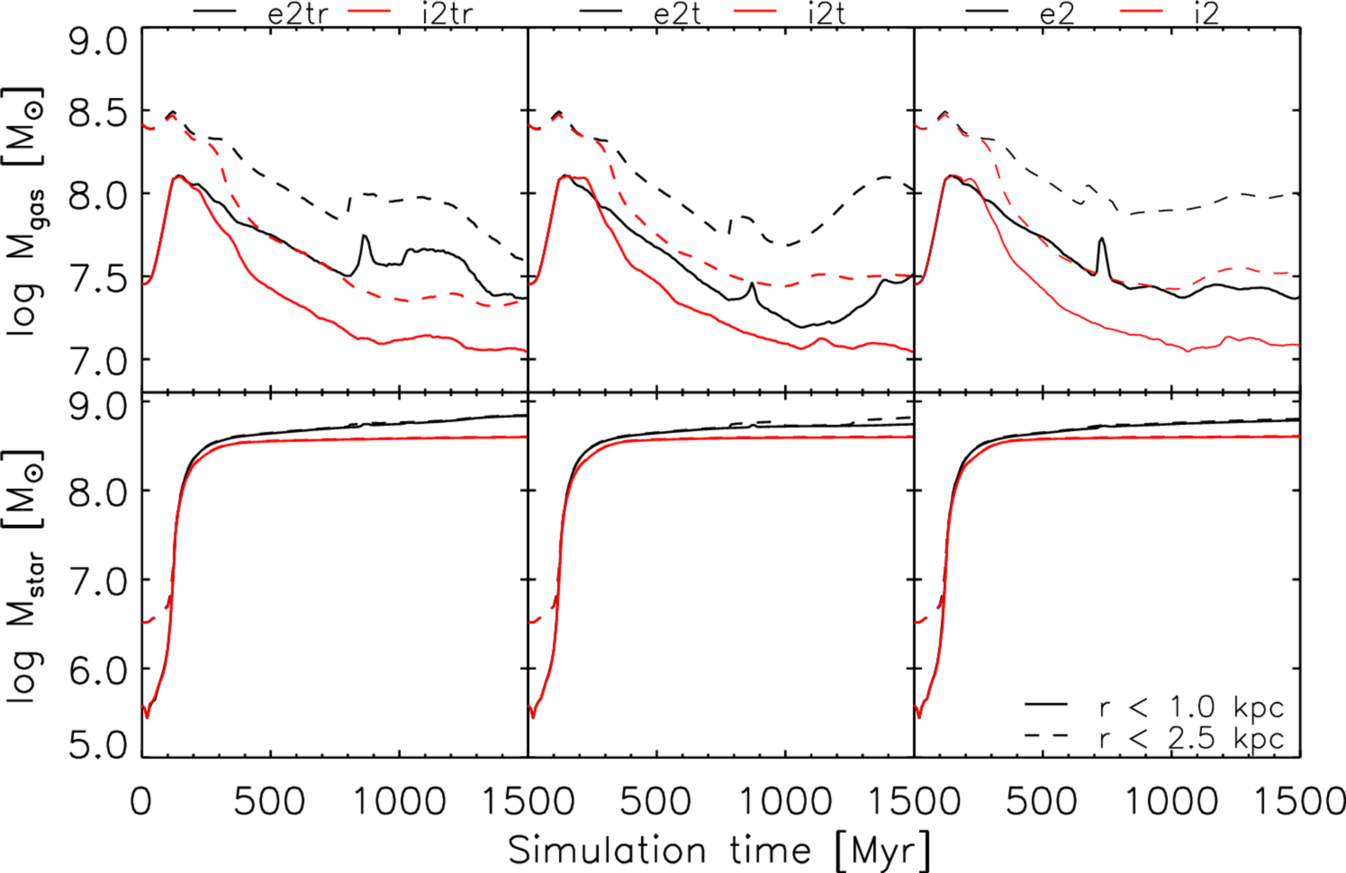} \hfill
 \includegraphics[width=\columnwidth]{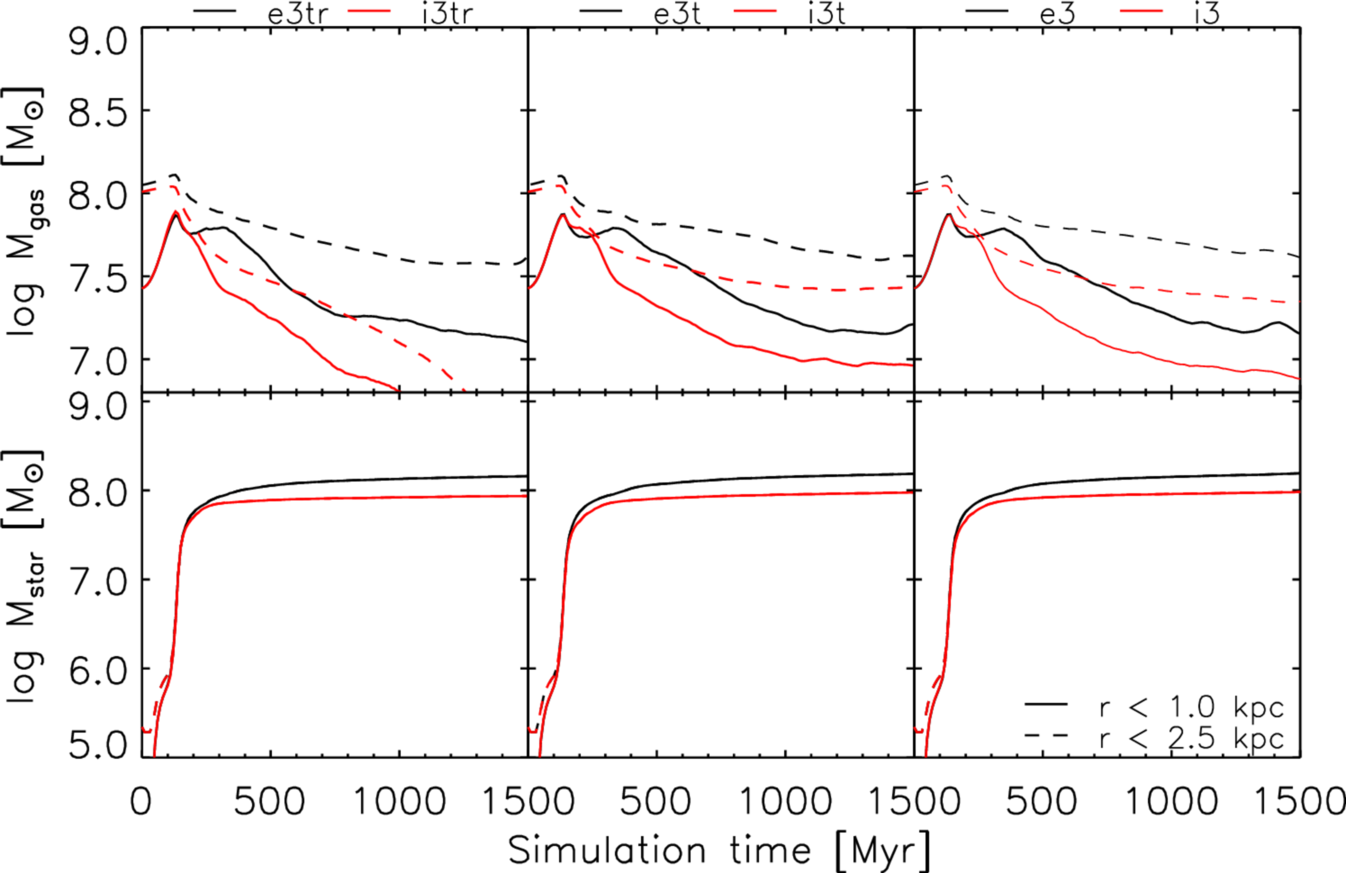} }
 \caption{Evolution of the gas (top) and stellar (bottom) mass within spheres of $r~=~1.0,~\mathrm{and}~2.5~\mathrm{kpc}$.
 In the left panel the embedded and isolated models of TDG candidate 2 are compared according to the included environmental
 effects as indicated above each column.
 The right panels show the same for TDG candidate 3.
 The line styles are indicated in the bottom right panel.
 }
 \label{fig:compare}
\end{figure*}

The embedded models generally double their stellar mass after a simulation time of $t_{sim}\approx300~\mathrm{Myr}$ until the end of the simulations. 
Starting from the same time, the stellar masses of the isolated models are approaching an equilibrium level with only a marginal increase in stellar mass until $1.5~\mathrm{Gyr}$
(see bottom panels of Figure~\ref{fig:compare}).

These characteristics show the importance of the tidal arm during the evolution of TDGs, as the tidal arm provides a rich reservoir of gas available for accretion and subsequent conversion into stars.
The accumulation of additional material allows the simulated TDGs to substantially increase their total mass within a radius of 2.5 kpc (see Figure~\ref{fig:mfrac25}~and~\ref{fig:mfrac25_s3}).

\section{Discussion}\label{sec:discussion}	

In order to explore the evolution of TDGs by consistently involving the impact of the tidal arm, we performed for the first time chemodynamical high resolution simulations of TDGs embedded into the tidal arm 
and compared these simulations to simulation runs of isolated TDGs, i.e. simulations with the same properties of the proto-TDG but neglecting the extended gas distribution of the tidal arm.
Moreover, the TDGs' initial structure is taken from a large scale merger simulation instead of starting form idealized isolated and symmetric models \citep[e.g.][]{ploe14,recchi15}.
As TDGs form and evolve in the vicinity of their parent galaxies, the influence of their environment has to be taken into account when studying this class of galaxies.
Therefore, we expose our TDG candidates to different combinations of environmental effects, such as the tidal field of the parent galaxies, 
and/or ram pressure caused by the motion through the CGM.

The evolution of the TDG candidates at test can be described by two principal phases.
At first, the initially already Jeans unstable proto-TDGs, with masses of $M_\mathrm{TDG}=2.60\times10^8$ and $1.12\times10^8\,\msol$ 
and Jeans masses of $M_\mathrm{J}=6.37\times10^7$ and $6.70\times10^7\,\msol$ 
within a radius of $2.5\,\mathrm{kpc}$, for candidate~2 and 3, respectively, rapidly collapse on the scale of a free-fall time. 
During this evolutionary stage a strong starburst occurs with a peak SFR of $\sim3.5\,\usfr$ for the models of candidate~2 and $\sim1.3\,\usfr$ for candidate~3.

After approximately $250-300\,\mathrm{Myr}$ the collapse subsides and the further evolution is determined by the amount of gas availability for accretion.
As the embedded models are able to refuel their gas reservoir, these models generally retain higher gas masses.
At this phase, the SFR follows nicely the available amount of gas, with a stronger decline for the isolated models (starvation) and 
an order of magnitude lower rates in the final simulation time compared to the embedded models.

This behaviour favours the formation of TDGs out of Jeans unstable gas clouds, as already described by \citet{elmegreen93}, 
rather then a clustering of stars in the tidal arm followed by gas accretion \citep[e.g.][]{barnes92}.
Nonetheless an unambiguous conclusion on the initial formation process of the proto-TDG might not be found from the presented work,
as the unstable cloud has been formed in the \citet{fouquet12} simulations at a timestep well before the initial data 
were extracted and therefore the details of its formation cannot be explored by the same detail.

At first sight, the peak SFRs in the order of $1~\usfr$ during the collapse of the proto-TDGs seem too high compared to observation, 
but are in agreement with the general evolution of TDGs as described by \citet{hunter00}.
Furthermore, these peak SFRs of our simulations confirm the findings of \citet{recchi07,ploe14,ploe15}, showing that TDGs can survive strong starbursts without being disrupted, despite their lack of DM.
The final self-regulated equilibrium SFRs in the range of $10^{-2}~\usfr$ agree well with observationally derived SFRs.

RPS is expected to be one of the main drivers of DG's morphological transformation, from gas-rich dwarf irregular galaxies to dEs, during the infall into galaxy clusters \citep[e.g.][]{lisker06,boselli14}.
Due to the absence of a supporting DM halo in TDGs, this class of dwarf galaxies should be particularly more vulnerable to RPS, 
as it has been demonstrated by \citet{smith13}. 
Thereby, they found that even the stellar component can be severely effected if the gas is removed fast enough and to a large gas fraction and to a large mass fraction.

However, young TDGs in the formation process and during their early evolution are embedded in a tidal arm consisting of gas and stars.
With its extended gas distribution it shields the TDG from the CGM, therefore, ram pressure is acting on the tidal arm and not directly on the TDG, at least as long as the arm is not completely dissolved.
Even after the arm has vanished, ram pressure does not act face-on onto the gas disk of the TDG. 
In a more realistic case, coplanar rotation of the TDG and the tidal arm must be assumed, so that ram pressure would hit the TDG edge-on.

Within the 15 TDG candidates of the original \citet{fouquet12} simulation the average angle between the TDGs orbit and internal rotation direction is $25^\circ\,\pm\,13^\circ$ 
and the orbital velocities are low with $v_\mathrm{orb}\,=\,148\,\pm\,40~\mathrm{km~s}^{-1}$ \citep[see][Table A.1]{ploe15}.
These numbers in combination with the presence of the tidal arm strengthen the presented results which suggest that RPS only has a minor influence on the early evolution of TDGs.

Moreover, the effect of RPS might be over-estimated by the presented simulations of the embedded models 
due to the truncation of the tidal arm at low densities during the initialisation of the simulation runs and the related adjustment of the CGM density. 
The assumption of a static CGM with respect to the merger remnant poses another source of uncertainty on the impact of ram pressure on the evolution of embedded TDGs.

As RPS and the tidal field only have minor influence on the early evolution of TDGs, the presence of the tidal arm has the largest impact on their juvenile evolution, 
i.e. after the initial collapse until the decoupling from the arm.
It provides a large reservoir of gas which can be accreted after the initial collapse and the related star burst.
The refuelling during the juvenile formation phase leads to an almost doubling of their stellar mass, and up to 3 times higher SFRs while still retaining higher gas masses and fraction 
throughout most of the simulation time, compared to the isolated models.

Under the assumption of a static CGM with respect to the merger remnant, the tidal arm further shields the TDGs from the CGM and protects them against 
RPS caused by the motion along their orbits around the host galaxies.
The extended gas distribution of the tidal arm has to dissolve or be removed before ram pressure can act on a young TDG directly, 
leading to a significant delay before RPS sets in. 
This allows TDGs to further accumulate matter, grow in mass and to stabilise themselves against the parents tidal field and RPS.
Within the presented simulation the delay time is of the order of a few $100~\mathrm{Myr}$, which most likely underestimates the true time scale of the decoupling of TDGs form their host tidal arm 
due to the truncation of the arms density distribution (for details see Section~\ref{sub:setup}).
Nevertheless it provides enough time to convert the majority of the TDGs initial gas content into stars before ram pressure can act on the TDG itself.

While we had already demonstrated by former numerical models that isolated TDGs, i.e. those detached from the gaseous tidal arm, can - though DM free - survive high SFRs and remain bound as dwarf galaxies. 
The new studies extend the TDGs to their environment as embedded TDGs and have shown that TDGs 

\begin{itemize}
 \item grow substantially in mass, due to the refueling along the tidal arm. 
 Depending on the structure, i.e. density and extent of the tidal arm, 
 TDGs can grow by more than one order of magnitude in mass within the central region and are able to double or triple their mass within a radius of $2.5~\mathrm{kpc}$. 
 
 \item are converting gas efficiently into stars at high SFRs. TDGs survive even strong starburst with SFRs of several $\usfr$. 
 
 \item are less sensitive to RPS than previous work \citep[e.g. by][]{smith13} suggests. 
 During their earliest evolution, the potentially devastating effect of RPS is strongly reduced by the shielding of the embedding tidal arm. 
 At later times, when the gaseous tidal arm has disappeared, RPS could act efficiently. 
 Due to the fast conversion into star-dominated systems the potential of being harmed by RPS is, however, 
 reduced because the lower gas fraction makes them less vulnerable to destruction by this gas-gas interactions. 
 
 \item remain bound and survive as star-dominated dwarf galaxies.
\end{itemize}

Although the models presented here are only a small sample of the numerous possible initial conditions, 
these models are representative for surviving TDGs which either escape form their parent galaxies or remain on bound orbits around them.
The selection of TDG candidates was taken under the aspect that they survived already in large-scale simulations \citep{yang14} because our modeling aims at zooming-in on survivor candidates and tracing their mass accretion and evolution with respect to the influence of the tidal field and ram pressure acting under realistic tidal-arm conditions. 
Therefore, it was out of our scope to derive an initial-to-final mass relations or any conclusions of the tidal-tail conditions on the TDG evolution.
This interesting aspect is not feasible for statistics from our few models but must be studied with an upcoming more comprehensive set of models.

\section*{Acknowledgements}
The authors thank Francois Hammer and Yanbin Yang for kindly providing their simulation data which served as initial data for our simulations.
The authors are grateful to Sylvia Ploeckinger for her advice to the numerical scheme and the referee Frederic Bournaud for his enthusiastic and supportive comments.
The software used in this work was in part developed by the DOE NNSA-ASC OASCR Flash Center at the University of Chicago.
The computational results presented have been achieved using the Vienna Scientific Cluster (VSC).




\bibliographystyle{mnras}
\bibliography{ref_r1} 








\bsp	
\label{lastpage}
\end{document}